\documentclass[12pt,preprint]{aastex}
\shorttitle{}
\shortauthors{Nesvorn\'y}
\begin{document}
\title{TOI-216: Resonant Constraints on Planet Migration}
\author{David Nesvorn\'y$^1$, Ond\v{r}ej Chrenko$^2$, Mario Flock$^3$}
\affil{(1) Department of Space Studies, Southwest Research Institute, 1050 Walnut St., \\Suite 300, 
Boulder, CO 80302, USA} 
\affil{(2) Institute of Astronomy, Charles University,\\ V Hole\v{s}ovi\v{c}k\'ach 2, CZ--18000 Prague 8, Czech Republic}
\affil{(3) Max-Planck-Institut f\"{u}r Astronomie, K\"{o}nigstuhl 17, 69117 Heidelberg, Germany}
\begin{abstract}
TOI-216 is a pair of close-in planets with orbits deep in the 2:1 mean motion resonance. The inner, Neptune-class 
planet (TOI-216b) is near 0.12 au (orbital period $P_{\rm b} \simeq 17$ d) and has a substantial orbital eccentricity 
($e_{\rm b} \simeq 0.16$), and large libration amplitude ($A_\psi \simeq 60^\circ$) in the resonance. The outer planet 
(TOI-216c) is a gas giant on a nearly circular orbit. We carry out $N$-body simulations of planet migration in a 
protoplanetary gas disk to explain the orbital configuration of TOI-216 planets. We find that TOI-216b's migration 
must have been halted near its current orbital radius to allow for a convergent migration of the two planets into the resonance. 
For the inferred damping-to-migration timescale ratio $\tau_e/\tau_a \simeq 0.02$, overstable librations in the resonance 
lead to a limit cycle with $A_\psi \simeq 80^\circ$ and $e_{\rm b}<0.1$. The system could have remained in this 
configuration for the greater part of the protoplanetary disk lifetime. If the gas disk was removed from inside out, this 
would have reduced the libration amplitude to $A_\psi \simeq 60^\circ$ and boosted $e_{\rm b}$ via the resonant interaction 
with TOI-216c. Our results suggest a relatively fast inner disk removal ($\sim 10^5$ yr). Another means of explaining the 
large libration amplitude is stochastic stirring from a (turbulent) gas disk. For that to work, overstable librations 
would need to be suppressed, $\tau_e/\tau_a \simeq 0.05$, and very strong turbulent stirring (or some other source of 
large stochastic forcing) would need to overcome the damping effects of gas. Hydrodynamical simulations can be 
performed to test these models.
\end{abstract}
\keywords{Exoplanets}
\section{Introduction}
TOI-216 hosts a pair of close-in exoplanets discovered by the \textit{Transiting Exoplanet Survey Satellite} (TESS) 
(Kipping et al. 2019; Dawson et al. 2019, 2021; Table 1). The initial analysis confirmed two planets -- an inner Neptune 
with a significant orbital eccentricity and an outer half-Jupiter -- with orbits near the 2:1 resonance. The physical 
characterization of the system was refined from stellar modeling, new transit observations, Radial Velocity (RV) 
and Transit Timing Variation (TTV) analysis. TOI-216 is a main sequence K-dwarf with the mass $M_* = 0.77$ 
$M_\odot$ and slightly sub-stellar metallicity. The key characteristics of the planetary system are: (i) $m_{\rm b}=0.059$ 
$M_{\rm Jup}$ and $m_{\rm c}=0.56$ $M_{\rm Jup}$ ($<5$\% 1$\sigma$ uncertainties), (ii) practically coplanar orbits, 
(iii) $e_{\rm b}=0.16$ and $e_{\rm c}<0.01$, and (iv) $P_{\rm b}=17.097$ d and $P_{\rm c}=34.552$ d. The two planets are 
firmly \textit{in} the 2:1 resonance, exhibiting resonant librations with the amplitude $A_\psi = 60 \pm 2$ deg 
(only 3\% uncertainty, Fig. \ref{res}; Dawson et al. 2021). The large resonant amplitude of the two planets in
the resonance is therefore very well established.

Having such an accurate characterization of a \textit{fully} resonant system of two exoplanets is rare. The classic GJ876 
RV-system consists of four known planets, three of which are in a chaotic Laplace resonance (1:2:4; Marcy et al. 2001;
Laughlin et al. 2005; Rivera et al. 2005, 2010; Nelson et al. 2016; Millholland et al. 2018). The Laplace resonance in 
GJ876 was assembled by planetary migration in a protoplanetary gas disk (Snellgrove et al. 2001, Lee \& Peale 2002, 
Crida et al. 2008, Delisle et al. 2012, Mart\'{\i} et al. 2013, Batygin et al. 2015, Dempsey \& Nelson 2018). Definitive 
inferences about the disk properties, however, are difficult to obtain in this case (Cimerman et al. 2018). 
Most \textit{Kepler} planets have non-resonant orbits. Kepler-9 is an iconic discovery of the \textit{Kepler} mission 
(Holman et al. 2010). Here the two planets -- now thoroughly characterized from TTVs (Freudenthal et al. 2018) -- are 
awkwardly placed near (but not in) the 2:1 resonance, in a domain where resonant librations exist ($\delta=3.1$ and 
$\Psi=5.7$ in Fig. \ref{res}B, but outside the plotted range), a configuration that suggests additional perturbations 
(e.g., scattering). 

Dawson et al. (2021) reported a preliminary dynamical analysis of TOI-216. They integrated a sample of best-fit 
solutions for 1 Myr and found them to be dynamically stable. Resonant capture was illustrated in two examples. 
In either of them, TOI-216c was migrated inward using the user-defined force in the \texttt{mercury6} integrator 
(Chambers et al. 1999). The radial migration of TOI-216b and eccentricity damping from disk torques were ignored. In the 
first example, TOI-216b was placed on an eccentric orbit before capture ($e_{\rm b}=0.08$; roughly a half of the 
present value), as needed for the planet to be captured with $A_\psi \sim 60^\circ$.  The resonant interaction with TOI-216c 
increased the orbital eccentricity of TOI-216b after capture. The migration torques on TOI-216c were then switched off, 
some $\sim 25,000$ yr after the start of the integration,
to obtain the final eccentricity $e_{\rm b} \simeq 0.16$. The second example in Dawson et al. (2021) invoked perturbations from an 
additional planet (approximated by an instantaneous change of TOI-216b's eccentricity vector).  Dawson et al. (2021) 
also suggested that the small mutual inclination of TOI-216b's and c's orbits ($i_{\rm M} \simeq 2^\circ$) 
could have been excited by the inclination 4:2 resonance. 
 
Here we develop a new dynamical model to explain the main characteristics of the TOI-216 
system (Sect. 2). Our model does not require special timing of the gas disk removal and/or additional planets. We 
suggest that TOI-216b was trapped near the transition to the inner, MRI-active zone of the 
protoplanetary disk (Flock et al. 2019), and waited for TOI-216c to migrate in (Sect. 3). The 2:1 resonance between 
the two planets could have been established relatively early during the disk lifetime (Sect. 4). Our model includes 
eccentricity damping (Sect. 5) and provides constraints on the migration-to-damping timescale ratio (Sect. 6). We 
invoke overstable librations to generate the large libration amplitude 
in the resonance (Sect. 7). A relatively fast inside-out disk removal is proposed to boost TOI-216b's 
eccentricity to the observed value (Sect. 8). This represents an interesting constraint on the physical mechanism 
of the inner disk dispersal. The influence of additional planets and turbulent stirring are discussed in Sects. 9 
and 10, respectively.     

\section{Resonant dynamics and TTVs}

\begin{figure}
\epsscale{0.49}
\plotone{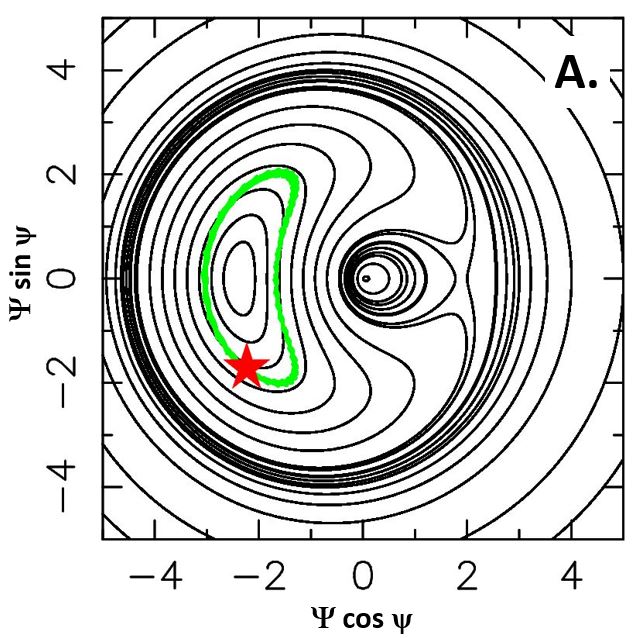}
\plotone{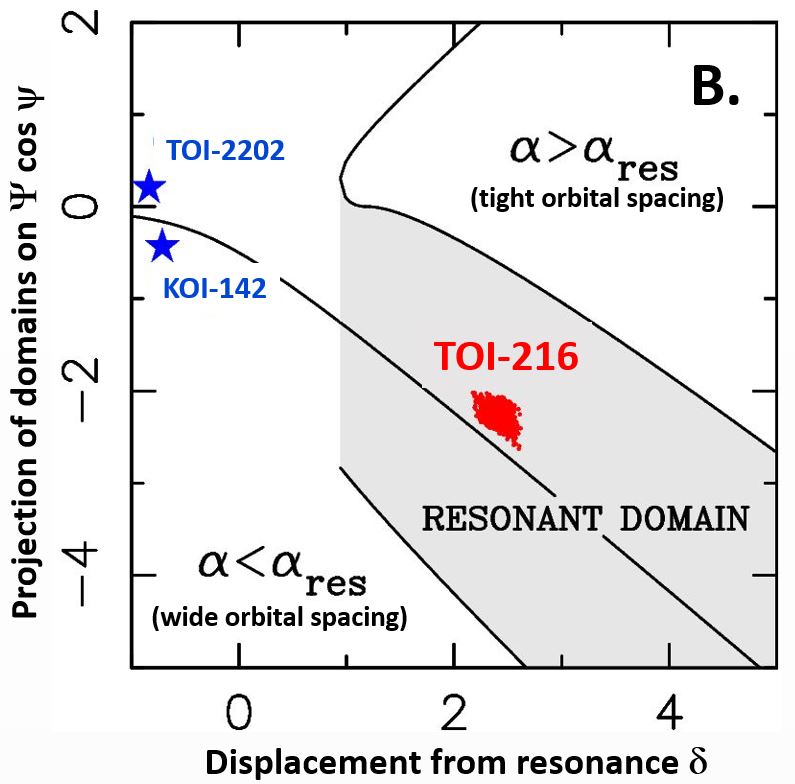}
\caption{A. The resonant librations of TOI-216 (green; the red star marks the current orbits) with the resonance portrait 
in the background (Nesvorn\'y \& Vokrouhlick\'y 2016). B. The resonant and near-resonant domains. TOI-216b,c are deep in 
the 2:1 resonance (the posteriori sample from Dawson et al. (2021) is shown as red dots). Exoplanet orbits are more commonly 
found just wide of resonances (e.g., KOI-142, TOI-2202). The orbital domain where the two orbits are just wide (or narrow) 
of the 2:1 resonance is indicated by $\alpha=a_{\rm 1}/a_{\rm 2} < \alpha_{\rm res} \simeq 0.63$ (or $\alpha > \alpha_{\rm res}$).
See Sect. 2. and Nesvorn\'y \& Vokrouhlick\'y (2016) for a definition of the resonant action $\Psi$, resonant angle $\psi$, 
and displacement from the resonance $\delta$.} 
\label{res}
\end{figure}

To highlight the orbital resonance in the TOI-216 system, it is useful to project the planetary orbits onto the 
representative plane defined by parameters $\delta$ and $\Psi \cos \psi$ (e.g., Nesvorn\'y \& Vokrouhlick\'y 2016; Fig. 
\ref{res}). The parameter $\delta$, defined in Eq. (23) in Nesvorn\'y \& Vokrouhlick\'y (2016) as a function of orbital 
elements of two planets, is approximately preserved by resonant dynamics. It can roughly be thought as a distance 
from the resonance. Away from the resonance, and if the orbital eccentricities are small, $\delta$ is related to the 
super-frequency (inverse of the usual super-period): it increases, in the absolute value, as the system moves away from the 
resonance. Negative values of $\delta$ imply that the planetary orbits are spaced more widely than the actual resonance 
($P_{\rm 2}/P_{\rm 1} > k/(k-1)$, where $P_{\rm 1}$ and $P_{\rm 2}$ are the orbital periods of the inner and outer planets, 
and $k=2$ for the 2:1 resonance). Large positive values of $\delta$ mean that the orbits are packed more tightly 
($P_{\rm 2}/P_{\rm 1} < k/(k-1)$).\footnote{Indices 1 and 2 are used to indicate parameters of the inner and outer planets, 
respectively. They are interchangeable with indices b and c wherever the text is specific to the TOI-216 system.}  

Resonant librations exist only for $\delta \geq \delta_* = (27/32)^{1/3} \simeq 0.945$ (Fig. \ref{res}B). 
Inside the libration island, $\delta$ is a measure of planet eccentricities at the equilibrium point around which 
the system librates. Action $\Psi$ and angle $\psi$ are resonant variables. Angle $\psi$ is a combination of the 
usual resonant angles $\sigma_1=2\lambda_2-\lambda_1-\varpi_1$ and $\sigma_2=2\lambda_2-\lambda_1-\varpi_2$, where
$\lambda_j$ and $\varpi_j$ are the mean and periapse longitudes of the two planets. Whereas the libration 
(circulation) of $\sigma_1$ and/or $\sigma_2$ is in general a good indicator of the resonant (non-resonant) configuration 
of orbits, exceptions are known to exist (e.g., Petit et al. 2020). It is therefore more definitive to verify on the 
behavior of $\psi$. 

The representative plane is the same for any first order resonance $k$:$(k-1)$ with $k \geq 2$; only the 
mapping from the orbital elements to $\delta$, $\Psi$ and $\psi$ depends of $k$. All planet pairs in and/or near the 
first-order resonances can therefore be placed on it. This makes the representative plane particularly useful for a 
comparative analysis of resonant and near-resonant systems. 

\begin{figure}
\epsscale{1.0}
\plotone{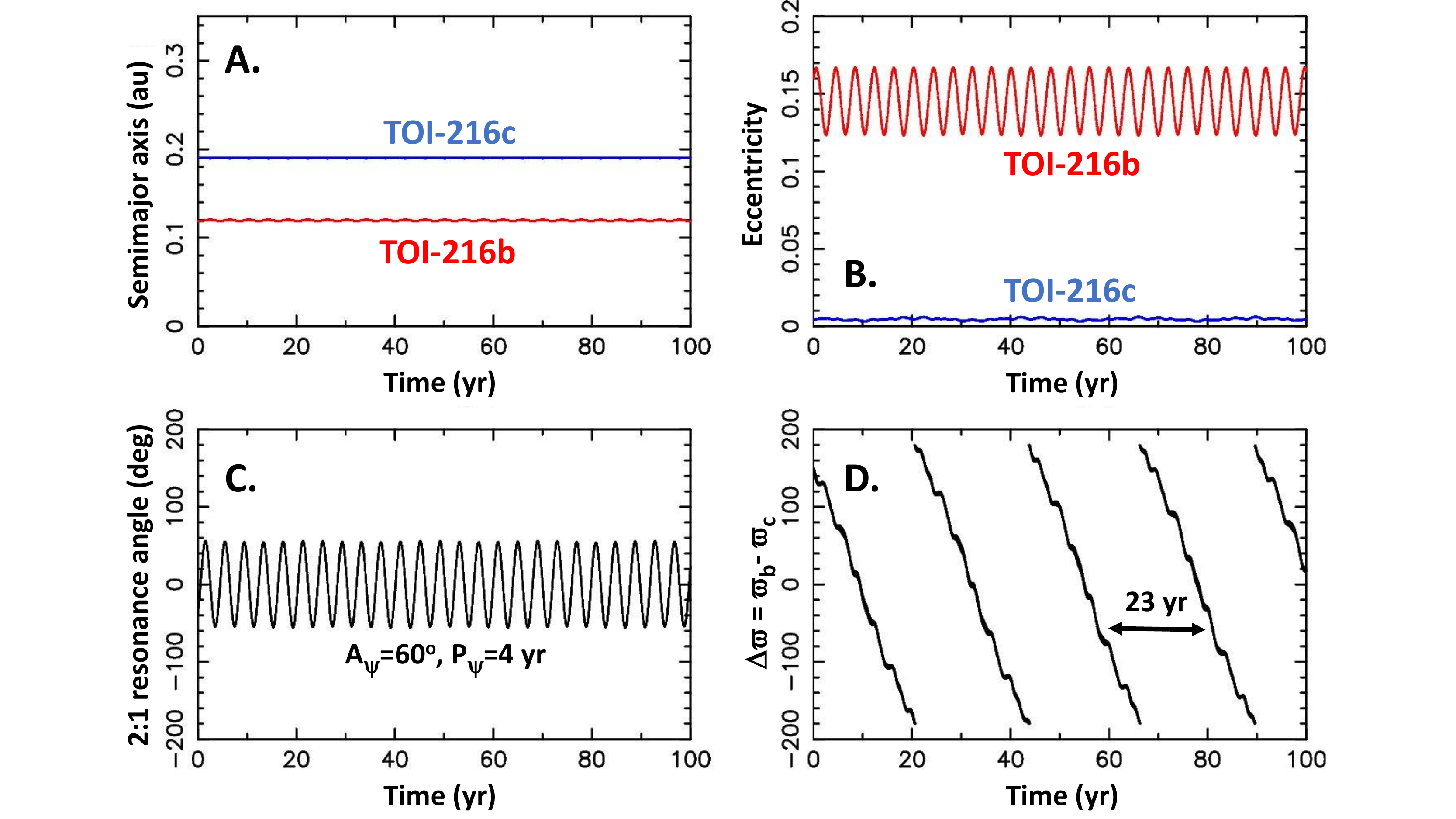}
\caption{The orbits of TOI-216 planets corresponding to the best-fit solution from Dawson et al. (2021). The resonant
librations are highlighted in panel C. In panel D, we show $\Delta \varpi = \varpi_{\rm b} - \varpi_{\rm c}$.} 
\label{real}
\end{figure}

The present orbits of TOI-216 planets have $\delta=2.1$, $\Psi=2.8$ and $\psi=220^\circ$, and this places them 
firmly in the 2:1 resonant island where $\psi$ librates around $180^\circ$ (Fig. \ref{res}). We performed a short 
integration of TOI-216 orbits with \texttt{swift\_mvs} (Levison \& Duncan 1994) starting from the best-fit parameters 
reported in Dawson et al. (2021). The resonant amplitude of $\psi$ is found to be $A_\psi \simeq 60^\circ$.
The period of resonant librations is $P_\psi \simeq 4$ yr. TOI-216b eccentricity shows resonant oscillations between 
0.124 and 0.166. The periapse longitude difference, $\Delta \varpi = \varpi_{\rm b}-\varpi_{\rm c}$, circulates 
in a retrograde sense with the period of $P_\varpi \simeq 23$~yr (Fig. \ref{real}). 

The resonant dynamics of TOI-216 planets explains the measured TTVs (Dawson et al. 2021). 
According to Eq. (2) in Nesvorn\'y \& Vokrouhlick\'y (2016), the main TTV period should be equal to the libration period
in the 2:1 resonance. From their Eq. (38), for $m_1 \ll m_2$, we have
\begin{equation}
P_{\rm TTV} \simeq P_1\; {P_\tau \over 2 \pi} \left( {m_2 \over M_*} \right)^{\!\!-2/3} 
\left[ {3 \over 2} f^2 \alpha_{\rm res}^2 \right]^{-1/3} \; , 
\end{equation}
where $P_1=17.1$ d, $P_\tau \simeq 3.3$ (Fig. 8 in Nesvorn\'y \& Vokrouhlick\'y 2016), $m_2/M_*=6.9 \times 10^{-4}$, 
$f=1.19$ ($f$ is one of the two standard resonant coefficients in the disturbing function expansion, here for 2:1),
and the resonant ratio of semimajor axes $\alpha_{\rm res}=a_1/a_2 \simeq 0.63$. This gives $P_{\rm TTV} \simeq 3.3$ yr, slightly 
shorter than the libration period measured from numerical integrations. The difference is caused by an approximation
in Nesvorn\'y \& Vokrouhlick\'y (2016) where only terms up to the first order in eccentricities are retained in the
disturbing function. The actual TTV measurements reported in Dawson et al. (2021) covered just under 900 d, some 60\%
of the full TTV/libration period. 

Similarly, for $m_1 \ll m_2$, the TTV amplitudes of the two planets are 
\begin{eqnarray}
A_1 & = & {P_1 \over 2 \pi} {P_\tau \over \pi} {A_\Psi \over k-1}\; , \nonumber \\
A_2 & = & {P_2 \over 2 \pi} {P_\tau \over \pi} {\sqrt{\alpha_{\rm res}} A_\Psi \over k-1} {m_1 \over m_2} \; ,
\end{eqnarray}
where $A_\Psi$ is the amplitude of the resonant oscillations of action $\Psi$ (zero for an exact resonance; Eq. (6) in 
Nesvorn\'y \& Vokrouhlick\'y 2016). For $A_\Psi \sim 0.7$ (Fig. \ref{res}A) this evaluates to $A_1 \sim 2.0$ d
and $A_2 \sim 0.35$ d, or $\sim 6,000$ min and $\sim 1,000$ min for the full TTV range over the whole libration cycle.
For comparison, the observed TTVs -- measured from the incomplete libration cycle -- are 4,000 min and 700 min, 
respectively (Dawson et al. 2021).

\section{Convergent migration}

The main goal of this work is to explain the orbital configuration of TOI-216 planets and obtain useful constraints
on the timescale of planet migration/damping and disk properties. Several conditions must be satisfied. First of all, 
the TOI-216 system almost certainly requires a {\it convergent} approach of planets into the 2:1 resonance. How the 
orbits converged is uncertain. The massive outer planet TOI-216c should have opened a deep gap in the gas disk 
(e.g., Crida et al. 2006) and slowly migrated inward by Type II (e.g., Kanagawa et al. 2018). The inner planet TOI-216b 
should have opened a shallower gap and migrated faster. In fact, the mass ratio of the two planets places them firmly 
in the regime of \textit{divergent} migration (Kanagawa \& Szuszkiewicz 2020). From this, we deduce that TOI-216b 
migration must have been stalled near its current $a_1 \simeq 0.12$ au to allow for TOI-216c to catch up.

There are several possibilities. TOI-216b could have stopped near the truncation of protoplanetary disk by its 
host star magnetosphere (e.g., Frank et al. 1992). Strong positive corotation torques are expected at the magnetospheric 
cavity radius (Masset et al. 2006, Romanova et al. 2019, Ataiee \& Kley 2021). The magnetospheric cavity radius
is estimated to be at $r_{\rm c} \sim 0.05$ au for $M_* = 1$ $M_\odot$, $R_* = 2$ $R_\odot$, $B_*= 1$ kG, and 
$\dot{M} = 10^{-8}$ $M_\odot$ yr$^{-1}$, where $B_*$ is the stellar magnetic field strength and $\dot{M}$ is the mass
accretion rate (Bouvier et al. 2014). The cavity is expected to expand as the mass accretion rate decreases (e.g., 
Liu et al. 2017). Alternatively, as we argue in this study, TOI-216b could have stopped at the transition radius from the 
outer dead zone (DZ) to the inner MRI-active zone (e.g., Kretke \& Lin 2012). We adopt a detailed model of the inner disk from 
Flock et al. (2019). The model matches interferometric observations of the inner edges of the disks around Herbig Ae/Be 
stars (Flock et al. 2016). In the nominal disk model for a solar type star (Flock et al. 2019), the zero-torque radius
-- which acts as a barrier for migrating planets -- is located at $\simeq 0.12$ au. 

The disk model was constructed in Flock et al. (2019) by solving for a hydrostatic equilibrium of a passively 
irradiated disk around an early solar type star. The turbulent gas viscosity $\nu$ was given by the usual $\alpha$ 
prescription: $\nu=\alpha c_{\rm s}^2/ \Omega$ (Shakura \& Sunyaev 1973), where $c_{\rm s}$ is the sound speed and $\Omega$ is the 
orbital frequency. The internal energy and radiation field were obtained as a steady-state solution of the equations 
describing heating, cooling and flux-limited diffusion of the disk's thermal radiation. The gas opacity was derived 
from Malygin et al. (2014). The dust opacities were computed for a mixture of astrophysical silicate and graphite 
from Wolf \& Voshchinnikov (2004). The model accounts for the sublimation of refractory grains close to the 
central star (Isella \& Natta 2005) and computes the dust-to-gas ratio as a function of temperature and radially integrated 
optical depth. When the temperature exceeds 
$T_{\rm MRI} \simeq 900$ K, the disk is assumed to be ionized and MRI active (Desch \& Turner 2015), with a boost to 
$\alpha$ ($=\alpha_{\rm MRI}$). The outer dead zone with $T<T_{\rm MRI}$ is characterized by a relatively low value 
of $\alpha$ ($=\alpha_{\rm DZ}$). 

Specifically, as a fiducial case for this study, we adopt the MREF disk from Flock et al. (2019). The MREF model uses the
stellar temperature $T_*=4300$ K, radius $R_*=2.6$ $R_\odot$ (as appropriate for a young solar-type star) and mass
$M_*=1.0$ $M_\odot$ to determine the stelar luminosity. The accretion-stress-to-pressure ratios, $\alpha_{\rm MRI}=0.01$
and $\alpha_{\rm DZ} = 0.001$, are taken to model the disk in the inner MRI-active region and outer dead zone, respectively.
The uniform mass accretion rate is set to $\dot{M} = 3.6 \times 10^{-9}$ $M_\odot$ yr$^{-1}$. The most important feature
of the MREF disk for this study is the surface density bump related to the change of $\alpha$ at the ionization transition
($T_{\rm MRI}$). The surface density bump produces a region of outward migration with a zero-torque radius on its
outer edge. The zero-torque radius position  mainly depends on the stellar luminosity.

We first consider the radial migration of TOI-216c. According to Kanagawa et al. (2015; also see Duffell \& 
MacFadyen 2013, Fung et al. 2014), a massive planet opens a gap with the contrast of perturbed and unperturbed 
gas densities
\begin{equation}
{\Sigma_{\rm gap} \over \Sigma} = {1 \over 1 + 0.04 K} 
\end{equation}
and 
\begin{equation}
K = { q^2 \over \alpha h^{5} }\; .  
\end{equation}
Here, $\Sigma_{\rm gap}$ is the surface density at the gap's bottom, $\Sigma$ is the unperturbed background surface 
density (i.e., before the planet is introduced), $q=m_{\rm p}/M_*$ is the 
planet-to-star mass ratio, and $h=H/r$ is the gas disk height-over-radius ratio. Numerically, for TOI-216c 
and Flock's MREF disk we have $q_{\rm c}=6.9 \times 10^{-4}$, $h=0.02$ (appropriate for the current orbital radius 
of TOI-216c, $r \simeq 0.19$ au), and $\alpha=10^{-3}$. This gives $K=1.5 \times 10^5$ and $\Sigma_{\rm gap}/\Sigma 
=1.7 \times 10^{-4}$. Clearly, TOI-216c is expected to open a deep gap in the gas disk and migrate inward in the 
type-II regime.      

Our understanding of type-II migration is incomplete (e.g., Robert et al. 2018, Chrenko \& Nesvorn\'y 2020), 
but here we use equations 
from Kanagawa et al. (2018) to get a rough sense of TOI-216c's migration timescale. The point is not to rigidly 
tie our expectations to these results -- and to the MREF disk from Flock et al. (2019) -- but to establish a 
theoretical reference in one specific case. In the linear theory, the disk torque exerted on a planet is 
given as a sum of Lindblad and corotation torques (e.g., Paardekooper et al. 2010). To a factor of the order of 
unity, the migration timescale of a deep-gap-opening planet can be approximated as  
\begin{equation}
\tau_a = -{a \over {\rm d}a/{\rm d}t} \sim 0.02 {1 \over \alpha h^3} {m_{\rm p} \over \Sigma r^2} {P \over 2 \pi}\ .
\label{taua}
\end{equation}
Note that $\tau_a$ is defined such that $\tau_a>0$ ($\tau_a<0$) for inward (outward) migration. 
For TOI-216c we have $\alpha=10^{-3}$, $h=0.02$, $\Sigma = 3,000$ g cm$^{-2}$ for $r=0.19$ au from MREF, 
$m_{\rm c} \simeq 0.56$ $M_{\rm Jup}$, $P \simeq 35$ d, and thus $\tau_{a,2} \sim 0.8$ Myr (we fixed $c=2$ in 
Eq. (19) in Kanagawa et al. 2018). This shows that TOI-216c is expected to migrate on a Myr-long timescale 
when it reaches $r=0.19$ au (the migration timescale would have been shorter at $r \gg 0.19$ au as 
$\tau_a \propto m_{\rm p} / \Sigma r^2$). The MREF disk with $\dot{M}=3.6 \times 10^{-9}$ $M_\odot$ yr$^{-1}$ 
has the surface density at 1 au that is a factor of $\sim2.5$ below the minimum mass solar nebula (MMSN, 
Hayashi 1981). If TOI-216c formed early and migrated in a more massive disk than MREF, its initial migration 
could have been faster. These initial stages are not, however, of an immediate concern here.  

We now turn our attention to TOI-216b and first discuss its migration in the MREF disk's dead zone ($r>0.15$ au). 
Adopting $q_{\rm b}=7.3 \times 10^{-5}$, $h=0.02$, $\alpha=10^{-3}$, we find $K \sim 1.7 \times 10^{-3}$, suggesting that 
TOI-216b should have also opened a relatively deep gap ($\Sigma_{\rm gap} / \Sigma \sim 0.014$). Since, according to Eq. 
(\ref{taua}), $\tau_a \propto M_{\rm p}$ in this regime, we find $\tau_{a,1} \sim 0.08$~Myr. The migration 
timescale of TOI-216b is therefore an order of magnitude shorter than that of TOI-216c: {\it the convergent migration 
cannot happen in this case}. The same result is obtained for a wide range of smooth disk models. For example, for 
higher viscosities and/or larger scale heights, TOI-216b would not be capable of opening a gap in the gas disk. 
It would therefore migrate even faster by full Type-I torques. We conclude that the convergent migration of 
TOI-216b and c, which is needed for their capture into the 2:1 resonance, is difficult to obtain in 
a smooth disk.    
 
Given the results discussed above, we find it likely that TOI-216b had arrived near its current orbital radius 
-- either formed there or migrated from larger orbital radii -- {\it before} the two planets were captured 
into the resonance. The radial migration of TOI-216b must have stopped near $r=0.12$ au to allow for TOI-216c 
to catch up. The transition between the outer dead and inner MRI-active zones in disks from Flock et al. (2019) 
offers a plausible mechanism for achieving this. Here the steep rise of the surface density with the orbital radius 
produces a planet trap (Masset et al. 2006), where planets can be held in place by very strong positive torques 
(exceeding, by at least an order of magnitude, the negative torques in the outer smooth disk; Flock et al. 2019;
see Schobert et al. 2019 for massive, viscously-heated disk models). TOI-216b could have been trapped relatively 
early during the disk lifetime and waited for TOI-216c to move in. 

In summary, the planet trap at the transition to the inner MRI-active disk region can generate favorable 
conditions for the convergent migration of TOI-216b and c, and their subsequent capture in the resonance.   

\section{Resonant capture}   

Once TOI-216b's migration stops (or at least considerably slows down), the two planets can be captured in the 2:1 
resonance. The capture is guaranteed if: (i) the evolution into the resonance is sufficiently slow (adiabatic),
and (ii) the orbital eccentricities are small. Mathematically, condition (i) requires that the time of resonance 
width crossing during migration, $\Delta t$, is longer than the libration period, $P_{\rm lib}$. Batygin (2015) 
computes
\begin{equation}
{P_{\rm lib} \over \Delta t} \sim 0.1 P_2 {\tau_1 - \tau_2 \over \tau_1 \tau_2} \left({M_* \over m_1 + m_2}\right)^{4/3} \ , 
\end{equation}
where $\tau_1$ and $\tau_2$ are the migration timescales of the two planets discussed in Sect. 3. For 
$P_{\rm lib}/ \Delta t < 1$, we therefore have that $\tau_2/P_2 >  0.1 / (q_{\rm b}+q_{\rm c})^{4/3} \sim 1.4 \times 10^3$ 
(here we assumed that $(\tau_1 - \tau_2) / \tau_1 \tau_2 \sim 1 / \tau_2$; i.e., TOI-216b is trapped and 
non-migrating), or $\tau_2 \gtrsim 100$ yr. This condition is easily satisfied for TOI-216c (we found $\tau_2 
\sim 0.8$ Myr in Sect. 3). 
We tested it numerically and found that the capture is guaranteed if $\tau_2 \geq 100$~yr, but can 
often happen even if $\tau_2 = 30$--100 yr when the pre-capture inner planet's eccentricity is very low ($e_1 \sim 
0.001$). If $e_1 \sim 0.05$, $\tau_2 = 50$ yr does not produce capture (even if the condition (ii) is satisfied). 

As for (ii), as the parameter $\delta$ crosses $\delta_*$ during migration (see Sect. 2), the area enclosed 
by the orbits in phase space of $\Psi$ \& $\psi$ (adiabatic invariant) must be smaller than the area enclosed by 
the separatrix (which forms at $\delta=\delta_*$; Henrard \& Lema\^{\i}tre 1983). For $m_1 \ll m_2$ and
$e_1 \gg e_2$, this leads to a simple condition 
\begin{equation}
e_1 < 1.6\; q_2^{1/3} \simeq 0.14
\end{equation}
for $q_2 = q_{\rm c}$ (Goldreich \& Schlichting 2014, Batygin 2015). This condition should be easily satisfied for TOI-216b 
if its orbital eccentricity is damped by gas. The stochastic forcing from turbulence in a gas disk is expected to 
excite $e_1 \sim 0.01$ (e.g., Nelson 2005). Once the resonant lock is established, it cannot be disrupted by turbulence (e.g., Adams et 
al. 2008, Rein 2012, Paardekooper et al. 2013, Batygin \& Adams 2017). 

\begin{figure}
\epsscale{1.0}
\plotone{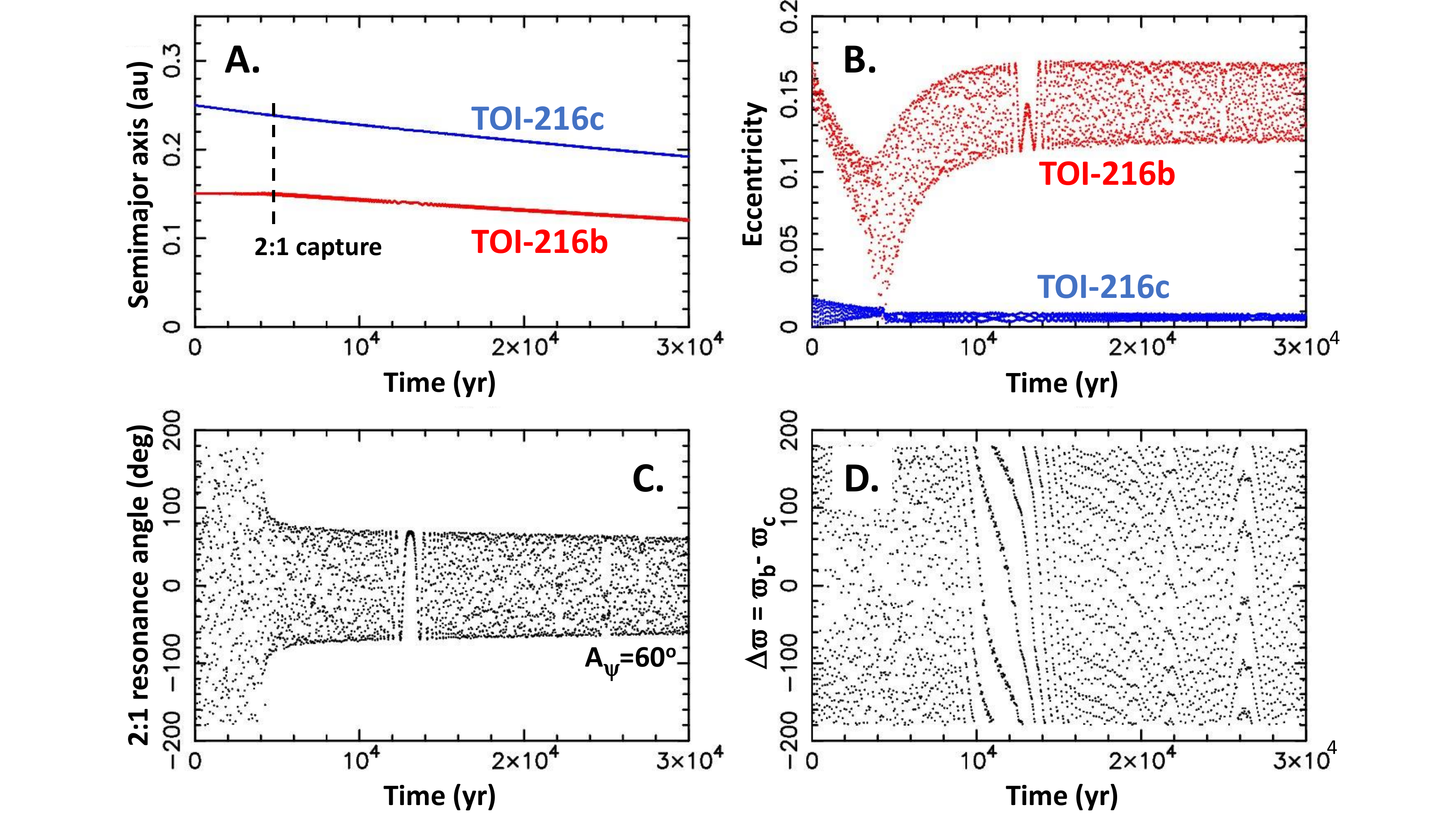}
\caption{Capture in the 2:1 resonance with a large pre-capture eccentricity of TOI-216b. Here we set $e_1=0.16$ 
at $t=0$ (panel B; similar to the present eccentricity of TOI-216b), $\tau_{a,2}=10^5$~yr, $\tau_{e,1}=5 \times 10^3$~yr, 
$p=0$ (Sect. 5), and let the two planets migrate into the resonance. For numerical convenience, the timescales used here 
are shorter, by a factor of $\sim 8$, than the ones expected for the MREF disk (Sect. 3). The time of capture is fine
tuned such that TOI-216b's eccentricity is not damped too much before the capture happens. This generates the initial libration 
amplitude similar to that of the real system (panel C). 
The simulation is stopped when $a_1 \simeq 0.12$ au and before the libration amplitude 
diminishes from damping effects. The special timing of capture -- just after $e_1$ is boosted by some process and before 
the disk is removed -- is difficult to avoid in this case.} 
\label{ex1}
\end{figure}

We verified by numerical integrations that, indeed, the capture is guaranteed if $e_1<0.14$. In fact, the above 
criterion is quite conservative and there is a good chance of capture even if the eccentricity is above the critical 
limit. If $e_1 \sim 0.1$ before capture, the capture into resonance leads to a large, but not too large, libration 
amplitude that would be compatible with measured $A_\psi \simeq 60^\circ$ (Dawson et al. 2021). With migration and 
\textit{without} the  eccentricity damping, which is the setup of numerical simulations performed here, 
the resonant coupling to TOI-216c would lead to a very high eccentricity of TOI-216b and escape from the resonance. 
Some eccentricity damping is therefore needed to stabilize the orbits in the resonance (Sects. 5 and 6). When the
eccentricity damping is included, however, it is difficult to explain why the inner planet's eccentricity could have
been so large before capture  (Fig. \ref{ex1}). Thus, invoking a large pre-capture eccentricity does not seem a
particularly attractive means of explaining the TOI-216 system.

We notice that -- if TOI-216c is placed beyond 0.25 au and migrated inward -- TOI-216b is very often captured in the 3:1 
resonance. The capture is nearly certain for $\tau_{a,2} \gtrsim 10^5$~yr and $e_b=0.001$--0.1. Moreover, if 
a sufficiently strong eccentricity damping is included, $e_{\rm b}$ reaches the equilibrium eccentricity (Sect. 6), and the 
TOI-216 planets end up surviving in the 3:1 resonance. Low eccentricities and short migration 
timescales could resolve this problem (e.g., $e_1=0.001$ and $\tau_{a,2} \lesssim 5 \times 10^4$ yr), but we argued for
$\tau_{a,2} > 10^5$ yr in Sect. 3. TOI-216c could have formed on a close-in orbit below the 3:1 resonance with TOI-216b, 
but this seems unlikely. Possibly the best solution to this problem hinges on the parametrization of the 
migration/damping torques. We numerically find that permanent 3:1-resonance capture only happens if the eccentricity 
damping is decoupled from the migration torques (the case with $p=0$ -- see the next section). The resonant capture 
can be avoided if the eccentricity damping happens at the constant angular momentum ($p=1$ in the next section). 
In that case, TOI-216b either avoids capture in the 3:1 resonance or is released from the resonance shortly after 
capture via a dynamical process related to overstable librations (Sect. 7).  

\section{Parametrization of disk torques}

The disk torques are usually implemented by adding the acceleration term
\begin{equation}
\mathbf{a}=-{\mathbf{v} \over 2 \tau_a} - 2 {{(\mathbf{v} \cdot \mathbf r}) \mathbf{r} \over r^2 \tau_e} - 
{v_z \over \tau_i} \mathbf{k}\ , \label{acc}
\end{equation}
where $\mathbf{r}$ and $\mathbf{v}$ are the position and velocity vectors, $v_z$ is the $z$ component of 
velocity, $\mathbf{k}$ is a unit vector along the $z$ direction, and $\tau_i$ is the inclination damping
timescale (not considered here). 
Using Lagrange equations, this can be written as
\begin{eqnarray}
{1 \over a} { {\rm d}a \over {\rm d}t }  & = & - {1 \over \tau_a} - {2 p e^2 \over \tau_e}\ , \label{da} \\
{1 \over e} { {\rm d}e \over {\rm d}t }  & = & - {1 \over \tau_e}\ . \label{de}  
\end{eqnarray}
with $p=1$. Subscripts 1 and 2 can be added here to indicate the inner and outer planets, respectively.  
The second term in Eq. (\ref{da}) appears because the radial force applied in the second term of Eq. (\ref{acc}) 
conserves the angular momentum $L=\mu \sqrt{GMa(1-e^2)}$, where $G$ is the gravitational constant, $M = M_*+m_{\rm p}$, 
and $\mu=M_* m_{\rm p} / M$. Thus, as the eccentricity is damped, the semimajor axis must decrease for $L = $ const. 

If $p=0$ instead, which is the case considered in the previous section, 
the semimajor axis migration would be independent of eccentricity damping. It is not clear which 
of the functional forms best expresses the actual disk torques - this is something that needs to be addressed 
by hydrodynamical simulations. Here we investigate cases with $0 \leq p \leq 1$. For $p<1$, we simply compute
the orbital elements of each planet at each time step, apply Eqs. (\ref{da}) and (\ref{de}), and return to 
the position and velocity vectors. 
 
\section{Equilibrium eccentricity}

In the 2:1 resonance, $e_{\rm b}$ is expected to grow toward the equilibrium eccentricity, $e_{\rm eq}$, that depends on 
the differential migration rate of the two planets, $1/\tau_a=1/\tau_{a,2} -1/\tau_{a,1}$, and the eccentricity damping 
rate, $1/\tau_e=1/\tau_{e,1}+(m_1/m_2)/\tau_{e,2}$, where $\tau_{e,j} = - e_j /({\rm d}e_j/{\rm d}t) $ 
(e.g., Lee \& Peale 2002, Goldreich \& Schlichting 2014, Deck \& Batygin 2015, Delisle et al. 2015, Terquem \& 
Papaloizou 2019; $\tau_{e,j}>0$ corresponds to eccentricity damping). The second term in the expression for 
$1/\tau_e$ can be neglected for TOI-216 as $m_1/m_2 \ll 1$. 

\begin{figure}
\epsscale{0.8}
\plotone{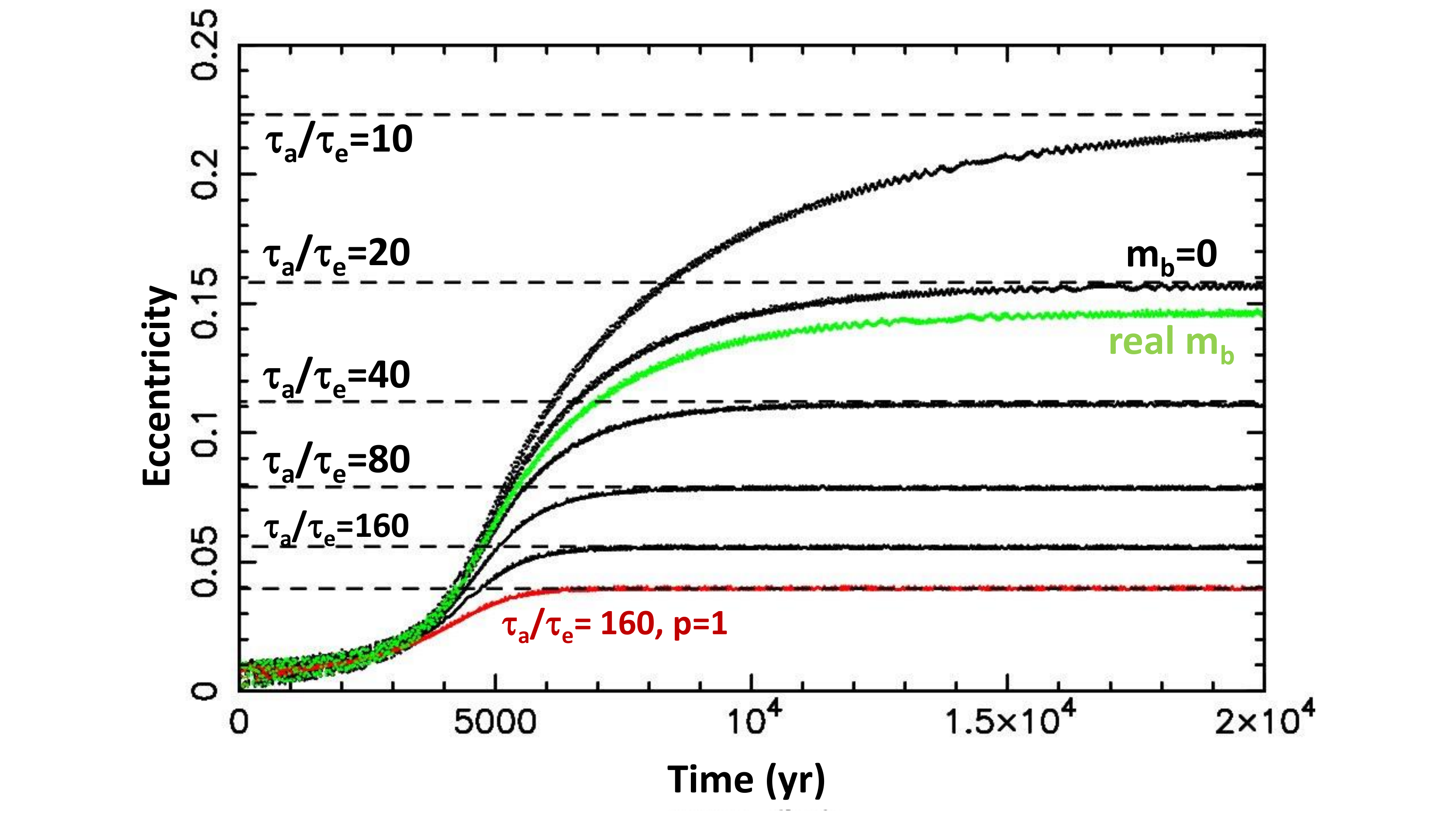}
\caption{The equilibrium eccentricities of TOI-216b inside the 2:1 resonance. Here we show the results for $p=0$ and $\tau_a/\tau_e=10$, 
20, 40, 80, and 160 (black lines from top to bottom). In all cases we set $m_1=0$ to be able to accurately compare the results with 
Eq. (11), plotted as dashed lines here. The green line is an example with the real mass of TOI-216b and $\tau_a/\tau_e=20$. The red line shows
the result for $p=1$, $\tau_a/\tau_e=160$ and $m_1=0$. The equilibrium eccentricity is a factor of $\sqrt{2}$ smaller than 
for $p=0$. This holds for any $\tau_a/\tau_e$. Thus, for example, the case with $\tau_a/\tau_e=40$ and $p=0$ is identical to 
the case with $\tau_a/\tau_e=20$ and $p=1$.} 
\label{eq1}
\end{figure}

By balancing the disk and resonant torques, and assuming $m_1 \ll m_2$, we find that
the inner planet, if captured in the 2:1 resonance, should approach the equilibrium eccentricity
\begin{equation}
e_{\rm eq} = \left( { 1 \over 2(1+p) } {\tau_e \over \tau_a} \right)^{1/2} \ 
\label{eqe}
\end{equation}
(e.g., Goldreich \& Schlichting 2014; $\tau_a>0$ and $\tau_e > 0$ assumed here; i.e., the convergent migration with 
damping). The equilibrium eccentricity thus mainly depends on
\begin{equation}
{\tau_e \over \tau_a} = \tau_{e,1} \left(  {1 \over \tau_{a,2}} - {1 \over \tau_{a,1}} \right) \ ,
\label{taus}
\end{equation} 
but there is also a dependence on $p$ (Fig. \ref{eq1}). For reference, to have $e_{\rm eq} \simeq e_{\rm b} = 0.16$, 
we would need $\tau_e / \tau_a \simeq 0.05$ for $p=0$ or $\tau_e / \tau_a \simeq 0.1$ for $p=1$ (i.e., the migration 
timescale would need to exceed the damping timescale by a factor $\sim 10$--20).   
 
\section{Overstable librations}

\begin{figure}
\epsscale{1.0}
\plotone{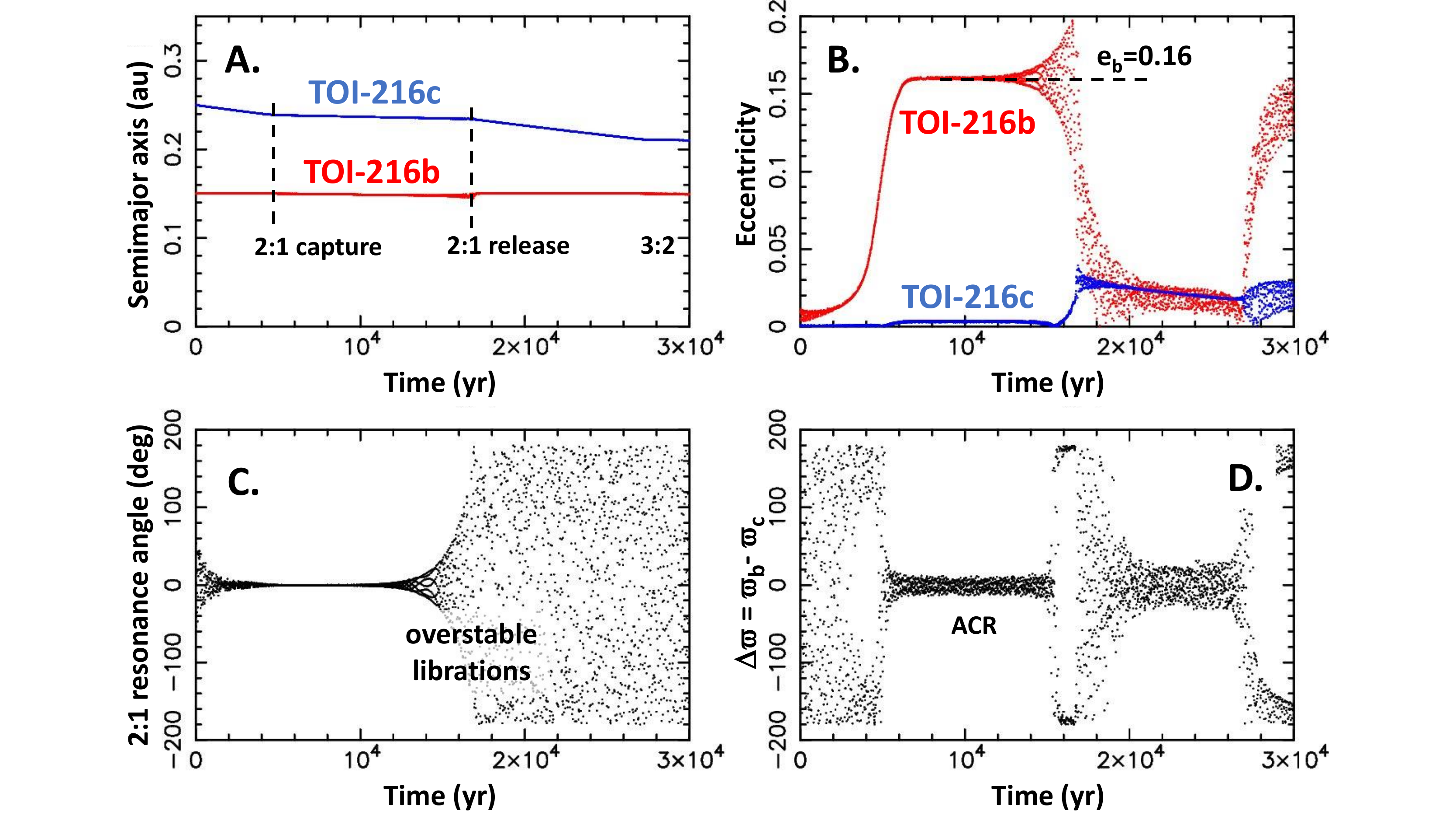}
\caption{Capture in the 2:1 resonance with a small pre-capture eccentricity of TOI-216b, and $p=1$. Here we set 
$e_1=0.01$ at $t=0$, $\tau_{a,2}=10^5$ yr, $\tau_{a,1}=-10^4$ yr for $a<0.15$ au, $\tau_{e,1}=10^3$ yr, and let the two planets 
migrate into the resonance. This leads to $e_1 \simeq 0.16$ shortly after capture, but the libration amplitude 
rapidly increases and the orbits escape from the 2:1 resonance at $t \simeq 16,500$ yr. ACR in panel D stands 
for the apsidal corotation resonance (Beaug\'e \& Ferraz-Mello 2003). For numerical convenience, 
here we use shorter migration/damping timescales than the ones expected for the MREF disk (Sect. 3).} 
\label{ex3}
\end{figure}

\begin{figure}
\epsscale{1.0}
\plotone{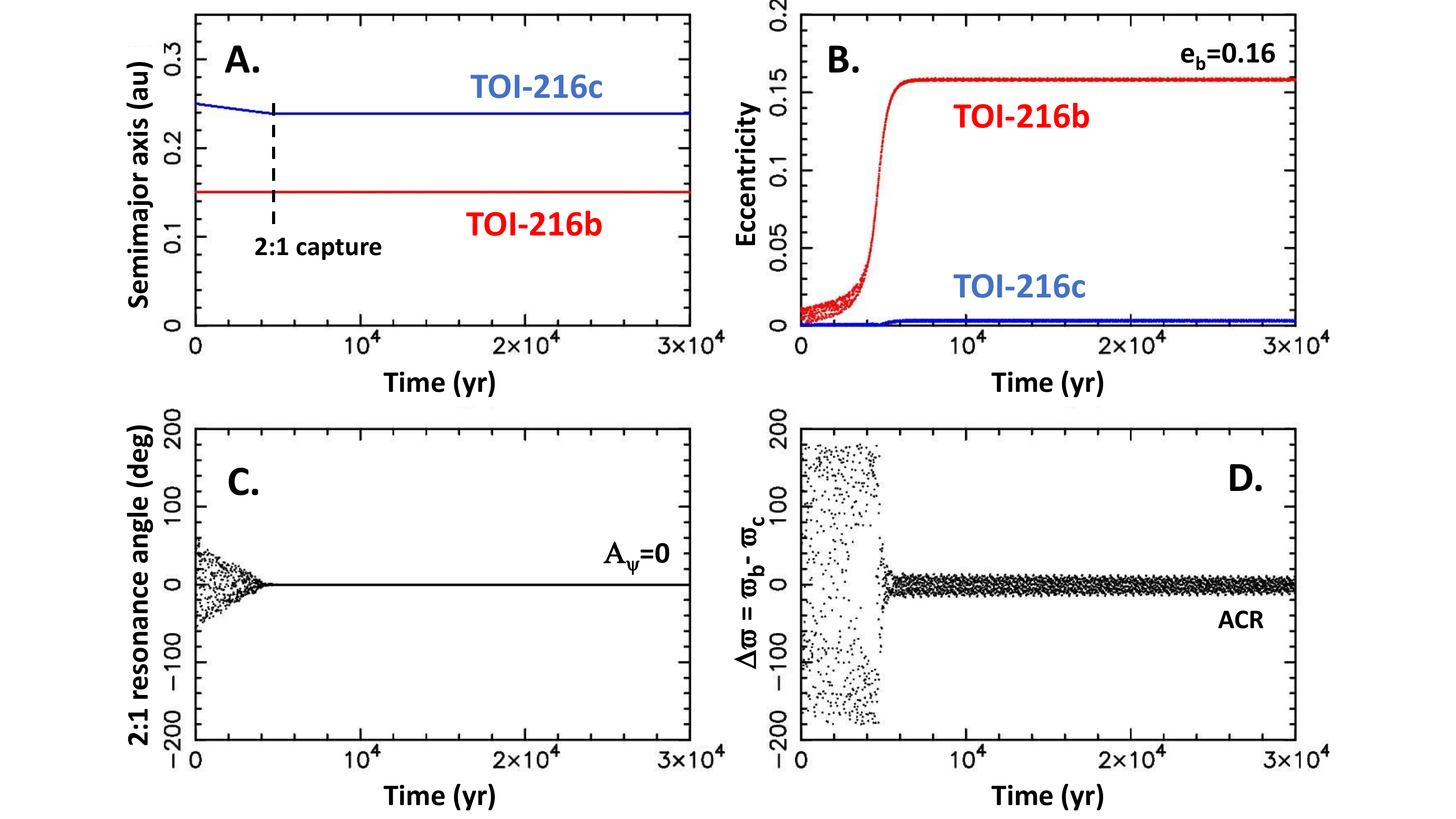}
\caption{Capture in the 2:1 resonance with a small pre-capture eccentricity of TOI-216b, and $p=0$. Here we set 
$e_1=0.01$ at $t=0$, $\tau_{a,2}=10^5$ yr, $\tau_{a,1}=-10^4$ yr for $a<0.15$ au, $\tau_{e,1}=500$~yr, and let the two planets 
migrate into the resonance. This eventually leads to $e_1 \simeq 0.16$ and a very small libration amplitude. The 
periapse longitudes end up in the apsidal corotation resonance with $\Delta \varpi \simeq 0$ (panel D; 
Beaug\'e \& Ferraz-Mello 2003). Short migration/damping timescales are used here for numerical convenience.} 
\label{ex2}
\end{figure}

Ideally, we would like to invoke a situation with $e_{\rm eq} \simeq e_{\rm b} = 0.16$. Unfortunately, for $p=1$, 
the resonant librations with such a large eccentricity are unstable. The libration amplitude would increase 
beyond limits and the planets escape from the resonance (the so-called overstable librations;
e.g., Goldreich \& Schlichting 2014, Deck \& Batygin 2015, Delisle et al. 2015). 
Our $N$-body integrations indicate that the escape timescale is extremely short once the overstable librations set in 
($<10^4$ yr; Fig. \ref{ex3}). One would therefore have to opportunistically remove the gas disk just at the right 
moment to end up with $A_\psi \simeq 60^\circ$. If $p=0$ instead, and assuming $e_{\rm b} \sim 0$ before capture, the libration 
amplitude is not excited (Fig. \ref{ex2}) -- the equilibrium point is stable in this case (Deck \& Batygin 2015).

The overstable librations is a dynamical effect that has been first noted in studies of dust particle dynamics in 
exterior resonances with the terrestrial planets (\v{S}idlichovsk\'y \& Nesvorn\'y 1994, Beaug\'e \& Ferraz-Mello 1994), 
and the tidal evolution of Saturn's satellites (Meyer \& Wisdom 2008). When Poynting-Robertson drag, tidal drag, or other 
generic drag/torque (Gomes 1995) is included in the equations of motion describing the resonant dynamics, the equilibrium 
point may become unstable. Whether the resonant equilibrium becomes unstable, and what the eventual fate of
a resonant system is, depends on a number of parameters. 

Goldreich \& Schlichting (2014) examined this problem for a pair of resonant planets with $m_1 \ll m_2$ and disk 
torques with $p \neq 0$. They found that the 2:1 resonant equilibrium is stable if 
\begin{equation}
e_{\rm eq} \leq \left ( {f q_2 \over 3 p} \right )^{1/3}  \; ,  
\end{equation}
where $e_{\rm eq}$ is related to the migration and damping timescales via Eq. (\ref{eqe}) (coefficient $f=1.19$ for 2:1;
Nesvorn\'y \& Vokrouhlick\'y 2016). For $q_2=q_{\rm b}=6.7 \times 
10^{-4}$ and $p=1$, this gives a condition $e_{\rm eq} \lesssim e_* = 0.06$. If $e_* < e_{\rm eq} < 2 e_*$ instead, the resonant
equilibrium is unstable, the libration amplitude increases, and the system eventually evolves onto a limit 
cycle with a fixed value of $A_\psi$. Finally, for $e_{\rm eq} > 2 e_* \simeq 0.12$, the resonant amplitude should 
increase beyond limits and planets escape from the resonance (Goldreich \& Schlichting 2014). 

The analytic thresholds are only approximate because $m_{\rm b} \neq 0$ (see Deck \& Batygin (2015) for a generalization 
of these formulas to massive inner planets). For TOI-216 and $p=1$, we numerically find that $e_{\rm eq}>0.09$ 
leads to escape from the resonance; this value is smaller than the analytic limit, $2 e_* \simeq 0.12$. In addition, 
whereas the analytic estimate suggests that the limit cycles should exist for $e_* < e_{\rm eq} < 2 e_*$ and {\it any} 
$p>0$, it is difficult to numerically find them when $p \rightarrow 0$. For example, 
$p<0.05$ would be needed for $e_* \sim e_{\rm b} =0.16$, but with the $p$ values this small, the transition from 
the stability to out-of-bounds growth of $A_\psi$ is razor sharp.   
 
\begin{figure}
\epsscale{1.0}
\plotone{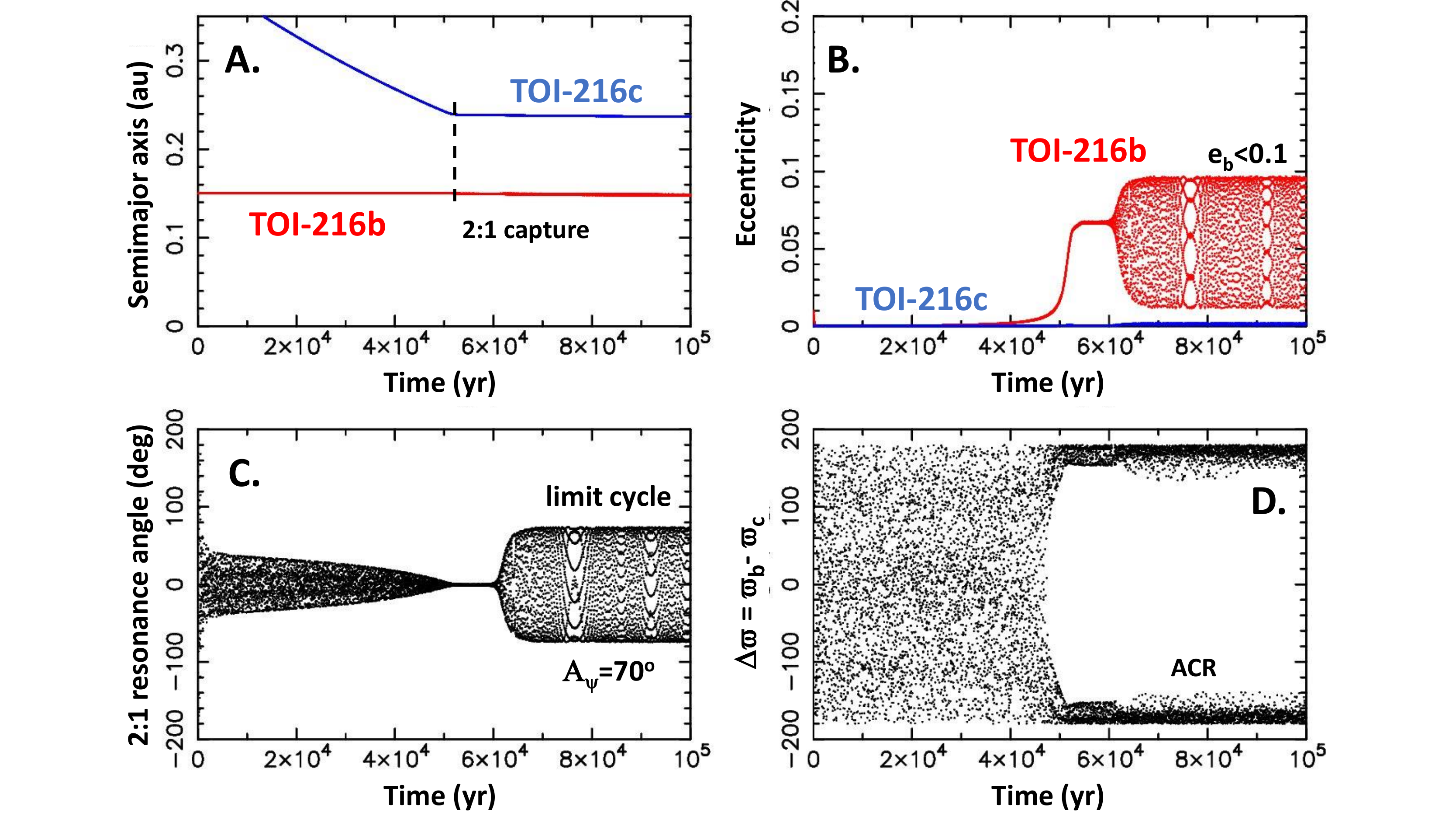}
\caption{Capture into the 2:1 resonance with a small pre-capture eccentricity of TOI-216b, and $p=1$. Here we set 
$e_1=0.01$ at $t=0$, $\tau_{a,2}=10^5$ yr, $\tau_{a,1}=-8,500$ yr for $a<0.15$ au, $\tau_{e,1}=150$~yr. This leads to 
overstable librations and a limit cycle with $e_1<0.1$ (panel B) and $A_\psi\simeq70^\circ$ (panel C). 
ACR in panel D stands for the apsidal corotation resonance (Beaug\'e \& Ferraz-Mello 2003).
Short migration/damping timescales are used here for numerical convenience.} 
\label{ex4}
\end{figure}

Figure \ref{ex4} shows an interesting case where the two TOI-216 planets are captured into the resonance with
$A_\psi \sim 0$ but then the overstable librations set in and lead to a limit cycle with $A_\psi \sim 70^\circ$.
The two planets could stay in this configuration almost indefinitely (as long as the disk conditions do not change). 
This is because TOI-216b is pushed in past the zero-torque radius where the strong positive torques on TOI-216b 
balance TOI-216c's migration torques (as long as the two planets are coupled via the 2:1 resonance). For that to 
happen, from the angular momentum conservation, we have that 
\begin{equation}
{\tau_{a,1} \over \tau_{a,2}} = - \sqrt{\alpha_{\rm res}} {m_1 \over m_2}
\label{mom}
\end{equation}
(weak eccentricity dependence neglected here). This gives $\tau_{a,1} / \tau_{a,2} \simeq - 0.084$.
We therefore set $\tau_{a,1} / \tau_{a,2} = - 0.084$ in Fig. 
\ref{ex4}. In reality, the system should have evolved to this equilibrium as the negative torque 
on TOI-216b increased when the planet was pushed farther past the zero-torque radius (Flock et al. 2019). We 
assume that the negative torques are strong enough such as TOI-216b cannot be pushed all the way into
the inner cavity.    

Whereas the model shown in Fig. \ref{ex4} could provide an explanation for the observed resonant amplitude of 
the TOI-216 planets, it falls short in matching TOI-216b's large eccentricity ($e_{\rm b} \simeq 0.16$). This 
leads to a question whether TOI-216b's eccentricity could have been excited when the gas disk was removed. Many 
additional simulations were performed here to investigate this problem. We tested the exponential disk dispersal 
at all radii, inside-out removal by the magnetospheric cavity expansion (Liu et al. 2017), and outside-in removal 
by photoevaporation (Alexander et al. 2014; as parametrized, e.g., in Ali-Dib \& Petrovich 2020). These simulations 
show that only the inside-out removal provides the desired effect: as the disk torques on TOI-216b cease but 
TOI-216c continues to migrate inward, $e_{\rm b}$ increases via the resonant interaction with TOI-216c. This is the 
base of the dynamical model described in the next section. 

\section{Putting things together}

\begin{figure}
\epsscale{1.0}
\plotone{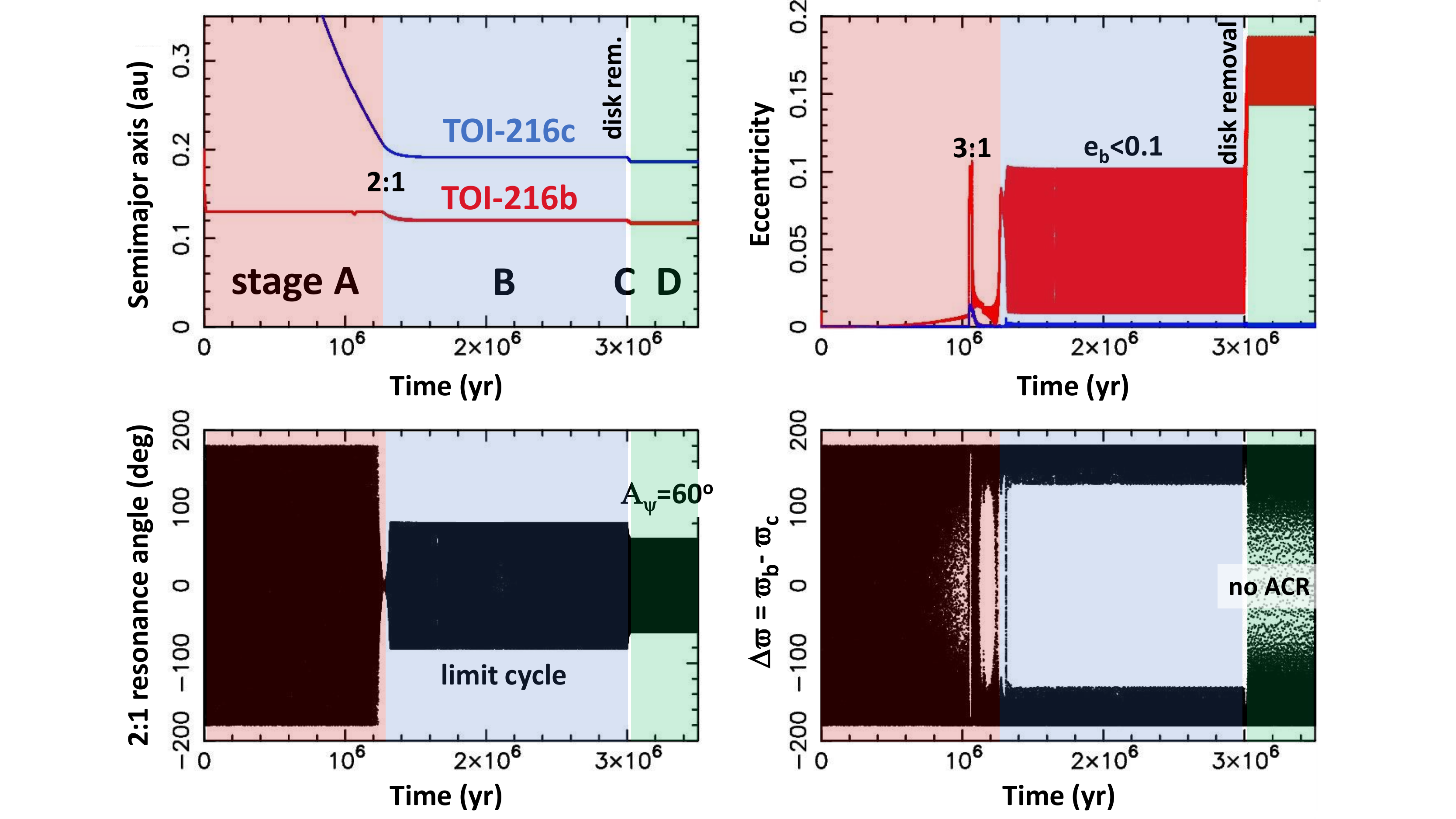}
\caption{The dynamical model for TOI-216 proposed in this work. The four stages, A--D, are described 
in the main text. Initially, TOI-216b (red line in panel A) was placed at 0.15 au and migrated to 
the zero-torque radius at 0.13 au ($t<0.05$ Myr), while TOI-216c continued migrating inward (blue line; stage A). 
The two planets become locked in the 2:1 resonance at time $t \simeq 1.25$ Myr after the start of the simulation, 
and overstable librations lead to a limit cycle with 
$e_{\rm 1} < 0.1$ (stage B). As the gas disk is removed from inside out just after $t=3$ Myr, damping of TOI-216b's 
eccentricity ceases, but TOI-216c continues to migrate in (stage C). The resonant coupling lifts b's $e$ to 
the observed value. The orbits of TOI-216b and c after the gas disk dispersal (stage D) are a good match 
to observations (cf. Fig. \ref{real}).} 
\label{model1}
\end{figure}

The proposed orbital history of TOI-216b,c is shown in Fig. \ref{model1}. There are several stages. In stage A
(time $t<1.27$ Myr), TOI-216b moves to the zero-torque radius where it is held in place. In the model shown in 
Fig. \ref{model1} we placed TOI-216b on an initial orbit with $a_{\rm 1}=0.15$ au and $e_{\rm 1}=0.01$ at let it
migrate in to $a_1=0.13$ au (this initial stage is difficult to see in the Fig. \ref{model1} because TOI-216b's
migration is relatively fast), but the details of this do not matter. TOI-216b could have formed much farther out
and reached 0.13 au after a long-range migration, or it could have formed at the pressure bump 
near the transition from the dead to MRI-active zones (Flock et al. 2019), where accumulating pebbles may trigger 
the streaming instability (Youdin \& Goodman 2005). TOI-216b's migration would have been short-range in the latter case
because the pressure bump and zero-torque radii are close to each other in the MREF disk (Flock et al. 2019). 
TOI-216c was placed at 1 au and migrated inward with $\tau_{a,2}=0.8$ Myr (Sect. 3) and $\tau_{e,2}=\tau_{a,2} / 50 
= 0.016$ Myr (the results are not sensitive to $\tau_{e,2}$).       

As for the torques applied to TOI-216b, the zero torque radius is placed at 0.13 au. For $a_{\rm 1}<0.13$ au, we linearly 
ramp up the positive torque such that $\tau_{a,1} / \tau_{a,2} = - 0.084$ at $r=0.12$~au. This is the expected 
orbital radius where the disk torques on TOI-216b and c should balance each other (Eq. \ref{mom}). For $a_{\rm b}>0.13$ 
au, we linearly increase the negative migration torque on TOI-216b until it reaches $\tau_{a,1}=0.08$ Myr, roughly 
at 0.15 au, which is the migration timescale inferred for TOI-216b in a smooth disk (Sect. 3). The eccentricity 
damping timescale is set to $\tau_{e,1}=0.02 |\tau_{a,1}|$ (i.e., the damping timescale scales linearly with the 
amplitude of $\tau_{a,1}$). The disk torques are parametrized by $p=1$ (Sect. 5) such that the overstable librations 
can arise.        

As TOI-216c migrates past $a \simeq 0.27$ au at $t\simeq1.07$ Myr, the two planets are briefly captured in the 
3:1 resonance. Overstable librations set in and the orbits are almost immediately released. That is one of 
the reasons why we prefer $p \neq 0$ (permanent capture in the 3:1 resonance would happen for $p=0$).
The two planets become locked in the 2:1 resonance at the beginning of stage B ($t\simeq1.27$ Myr). As TOI-216c 
continues migrating inward, it pushes TOI-216b past the zero-torque radius into the region with strong positive 
migration torques. This leads to $a_1 \simeq 0.12$ au, as anticipated from the design described above. Overstable 
librations set in immediately after capture and lead to a limit cycle with a large libration amplitude 
($A_\psi \simeq 80^\circ$), and $e_{\rm b} \lesssim 0.1$. This stage could last indefinitely as long as the disk conditions 
do not change; hence, no special timing needed. Here we assume that stage B lasts till $t=3$~Myr, a time interval comparable to the typical 
disk lifetimes (2--5 Myr; Williams \& Cieza 2011).  

Then, during stage C, the gas disk is removed from inside out (e.g., when the magnetospheric cavity expands; 
Liu et al. (2017); rebound not included). The principal disk torques on TOI-216b cease when the cavity moves past 
$\simeq 0.12$ au, but TOI-216c continues migrating in. This boosts $e_{\rm b}$ during stage C. The initial 
($e_{\rm I}$; at the beginning of stage C) and final ($e_{\rm F}$; at the end of stage C) eccentricities of TOI-216b 
are related by 
\begin{equation}
e_{\rm F}^2 \simeq e_{\rm I}^2 + \ln \left( 1 + {\Delta t \over \tau_{a,2}} \right)\; ,  
\end{equation}
where $\Delta t$ is the duration of stage C (i.e., the time between disk removal at 0.12 au and 0.19 au; e.g., 
Malhotra 1995). With $e_{\rm I}=0.05$, $e_{\rm F}=0.16$, $\tau_{a,2}=0.8$ Myr, we solve for $\Delta t$ to 
estimate $\Delta t \simeq 25,000$ yr, which is the duration of stage C used in Fig. \ref{model1}. 

Both TOI-216b and c migrate inward during stage C, but given its short duration these changes are difficult
to resolve on the scale of Fig. \ref{model1} (TOI-216b moves from $a_1 \simeq 0.12$ au to $a_1 \simeq 0.118$ au).  
The libration amplitude slightly decreases (as the adiabatic invariant is conserved; Henrard \& Lema\^{\i}tre 1983).
Finally, in stage D ($t>3.025$ Myr), the disk cavity moves beyond 0.2 au, and the orbital configuration freezes 
in place. The semimajor axes of the two planets are $a_1 \simeq 0.118$ au $a_2 \simeq 0.188$ au, the 
libration amplitude is $A_\psi \simeq 60^\circ$, the orbital eccentricity of TOI-216b oscillates between 0.14 
and 0.19. The final orbits are an excellent match to observations (Dawson et al. 2021). 

\section{Were there more inner planets?}

The model presented here has several advantages. It does not require additional planets. TTV and RV measurements do not 
reveal additional massive planets in the TOI-216 system, but the present limits are not too strong (Dawson et 
al. 2021). Hypothetical additional planets could have also been eliminated. Here we look into this possibility by 
investigating {\it Kepler}-class systems of super-Earths with $n_{\rm p}=2$, 3 and 4 inner planets. For simplicity, the inner 
planets are given masses $m_{\rm p}=m_{\rm b}/n_{\rm p}$; that is $m_{\rm p} \simeq 9.4$, 6.3, 4.7 $M_{\rm Earth}$ for $n_{\rm p}=2$, 3 and 
4, respectively. They are initially placed on nearly circular and nearly coplanar orbits between 
0.15 and 0.5 au, and migrate inward to produce resonant chains. The migration and damping torques are defined following 
the method described in the previous section. All collisions are assumed to be accretional. TOI-216c is placed on an 
outer orbit and migrated inward with $\tau_{a,2}=0.8$ Myr. 

Figure \ref{model2} shows a typical case for the system starting with three inner planets. After a series of 
resonant captures and instabilities, the three inner planets merge into a final planet with the mass identical to that 
of TOI-216b. TOI-216b and TOI-216c are then captured in the 2:1 resonance where overstable librations set in and increase 
the resonant
amplitude to $\simeq80^\circ$. TOI-216b's eccentricity shows oscillations reaching up to $e\sim 0.1$ during this stage.
The disk is removed after $t=3$ Myr with $\Delta t=25,000$ yr (the same as in the previous section) producing an eccentricity surge very
similar to that shown in Fig. \ref{model1}. The final architecture of the model system is again a good match to TOI-216. 

The cases with $n_{\rm p}=2$ and $n_{\rm p}=4$ produce the same dynamics with the inner planets merging and eventually 
forming a single, more massive inner planet, which is then captured into the 2:1 resonance with TOI-216c. The final
configuration of orbits is practically the same, independently of whether we start with one (previous section) or
$n_{\rm p}=2$--4 inner planets. This demonstrates that TOI-216 may have initially hosted a {\it Kepler}-like system of super-Earth/mini-Neptunes. 
The system would have been destabilized when TOI-216c moved in, and the inner planets would have merged into what is now TOI-216b. 
The initial number of planets, their masses and orbits, are unknown, and our only constraint is that the total mass of the 
inner system was $\sim 18.8$ $M_{\rm Earth}$, slightly exceeding that of Neptune.
The TOI-216 system would be special because massive TOI-216c formed and migrated all the way from its birth location to 
$a < 0.2$ au, whereas no such massive close-in planets affected the {\it Kepler} systems in general. 

\section{Stochastic forcing}

Here we investigate additional means of explaining the orbital configuration of TOI-216 planets. The turbulent stirring 
has been previously suggested to interfere with the capability of mean motion resonances to capture and retain {\it Kepler}-class 
planets in the resonant chains (e.g., Adams et al. 2008, Rein \& Papaloizou 2009, Rein 2012, Paardekooper et al. 2013, 
Batygin \& Adams 2017). The turbulent stirring arises when fluctuations within a turbulent disk produce a random 
gravitational field which then affects the embedded planets (e.g., Laughlin et al. 2004, Nelson 2005). Here we adopt 
the analytic formulation from Okuzumi \& Ormel (2013). In the limit of ideal MRI turbulence, the diffusion coefficients 
for $a$ and $e$ are estimated as 
\begin{eqnarray}
D_a & \simeq & 5.5\times10^{-4} \left( {\alpha \over 10^{-2} } \right) \left( {\Sigma a^2 \over M_* } \right)^2 a^2 \Omega \label{difa}\\ 
D_e & \simeq & 2.4\times10^{-3} \left( {\alpha \over 10^{-2} } \right) \left( {\Sigma a^2 \over M_* } \right)^2 \Omega \ . \label{dife} 
\end{eqnarray}
For our fiducial MREF disk, this evaluates to $D_a \simeq 3.5 \times 
10^{-13}$ au$^2$ yr$^{-1}$ and $D_e \simeq 1.3 \times 10^{-11}$ yr$^{-1}$ for TOI-216b and $D_a \simeq 2.9 \times 10^{-13}$
au$^2$ yr$^{-1}$ and $D_e \simeq 4.0 \times 10^{-12}$ yr$^{-1}$ for TOI-216c. We adopted $\Sigma=3,000$ g cm$^{-2}$, 
$\alpha=0.01$ for TOI-216b and $\alpha=0.001$ for TOI-216c. This choice is somewhat arbitrary because the $\alpha$
parameter changes with radial distance in our fiducial disk, and it is therefore not clear what effective $\alpha$ 
value should be used in Eqs. (\ref{difa}) and (\ref{dife}).  

The stochastic forcing was implemented in the $N$-body integrator. We apply stochastic kicks in $a$ and $e$ once every 
$\Delta t$, where $\Delta t < P_1, P_2$, and adjust the amplitude of these changes such that the random walk in $a$ and $e$
-- without damping/migration and mutual interaction between planets -- reproduces the expected characteristics of 
diffusion (with $D_a$ and $D_e$ given above). This should at least qualitatively replicate the diffusive evolution expected 
from the turbulent stirring. To separate the effects of turbulent stirring from those of overstability, all integrations 
reported here use $p=0$ (Sect. 5). In this case, the resonance equilibrium is stable and the resonant amplitude, if excited by
turbulence or other means, is expected to be reduced by the damping effects of gas. This means that the resonant 
amplitude can become large only if the turbulent stirring is strong enough to overcome the amplitude damping. 
We set $\tau_e/\tau_a=0.05$ to generate $e_1 \simeq 0.16$ via resonant interaction.  

With the nominal diffusion coefficients reported above, the orbital evolution of the two planets is very 
similar to that without any stirring (e.g., Fig. \ref{ex2}). If TOI-216c starts beyond the 3:1 resonance with TOI-216b, the 
two planets become captured in the 3:1 resonance with $A_\psi \simeq 0$ and remain in the resonance. If TOI-216c starts 
closer-in and does not cross the 3:1 resonance during migration, the two planets are captured in the 2:1 resonance with 
$A_\psi \simeq 0$. The nominal turbulent stirring is clearly not strong enough to produce any significant
libration amplitude. We therefore test models in which the magnitude of stochastic forcing is increased. We believe that 
this is reasonable because the existing models of stochastic forcing in a protoplanetary disk are approximate. 
Hydrodynamical instabilities can generate large scale motions of gas which cannot be accurately represented with the 
Shakura-Sunayev $\alpha$ prescription (e.g., Flock et al. 2017). For example, the Rossby instability at the inner edge of 
the dead zone can generate a vortex that would interact with TOI-216b. 

Any detailed modeling of these effects is left for future work. Here we just ask, and answer, the question of how much 
stochastic forcing is needed to produce $A_\psi \sim 60^\circ$. We use the numerical scheme described above and increase the diffusion 
coefficients by a constant factor, $f>1$. Figure \ref{model4} shows the result for $f=400$. In this case, the two planets are 
captured in the 3:1 resonance for only a brief interval of time (near $t=1.05$ Myr). The subsequent capture in the 2:1 
resonance is permanent. The libration immediately reaches $A_\psi > 40^\circ$ due to the stochastic forcing and remains large 
during the whole disk lifetime. The disk is instantaneously removed at $t=3$ Myr after the start of the simulation; the 
timescale or means of disk removal do not matter in this case. The final orbits of TOI-216b and c are a good match to 
observations, featuring $e_1 \simeq 0.16$ and $A_\psi \simeq 60^\circ$. By testing different values of $f$ we find that 
$f>100$ is needed to avoid the permanent capture in the 3:1 resonance {\it and} have a reasonable chance to end up with 
$A_\psi \simeq 60^\circ$ in the 2:1 resonance. 
 
It is not clear whether such a large boost to the nominal stochastic forcing can be justified. The effects of turbulence,
as described in Okuzumi \& Ormel (2013), are relatively weak in the MREF disk. To obtain $f > 100$, $\alpha$ would have 
to be increased by a factor of $>100$ or $\Sigma$ would have to be increased by a factor of $> 10$ (Eqs. \ref{difa}
and \ref{dife}). None of these parameter choices is reasonable. If anything, the non-ideal MRI effects and/or gap 
opening would reduce -- not increase -- the strength of turbulent stirring experienced by the two planets. TOI-216b's interaction 
with an inner vortex could be more promising (e.g., Faure \& Nelson 2016). 
By switching different components of stochastic forcing on and off, we find 
that the diffusion in the semimajor axis of TOI-216b is mainly responsible for generating the large libration amplitude. 
Thus, the constraint obtained here can be formulated as $D_a > 3.5 \times 10^{-11}$ au$^2$ yr$^{-1}$ for TOI-216b. 
Hydrodynamical simulation can be used to test whether such a large stochastic forcing is possible.\footnote{Alternatively, 
one could consider a case with very weak amplitude damping (i.e., large $\tau_e$), but that would probably imply, 
with $\tau_e/\tau_a=0.05$, unrealistically long migration timescales.}   

\section{Discussion}

We find that $\tau_e/\tau_a \simeq 0.02$ for $p=1$ (Sect. 8) or $\tau_e/\tau_a \simeq 0.05$ for $p=0$ (Sect. 10).
From Eqs. (\ref{taus}) and (\ref{mom}), and neglecting factors $\sim 1$, it can shown that this is equivalent to 
$\tau_{e,1} / \tau_{a,1} \simeq 0.02$ or 0.05 in a smooth disk. In other words, the damping-to-migration timescale ratio 
is of the order of, or larger than, the disk aspect ratio ($h=0.02$) in the reference protoplanetary disk used here 
(Flock et al. 2019). This constraint is obtained from analyzing the TOI-216 system in a regime where $e>h$. In the 
low-eccentricity regime with $e<h$, it is found from theory and hydrodynamical simulations that $\tau_e/\tau_a \sim h^2$ 
(Papaloizou \& Larwood 2000, Cresswell et al. 2007). If this is applied to the TOI-216 system, it would be difficult 
to understand how the very strong eccentricity damping was overcome to generate $e>h$. 

\begin{figure}
\epsscale{1.0}
\plotone{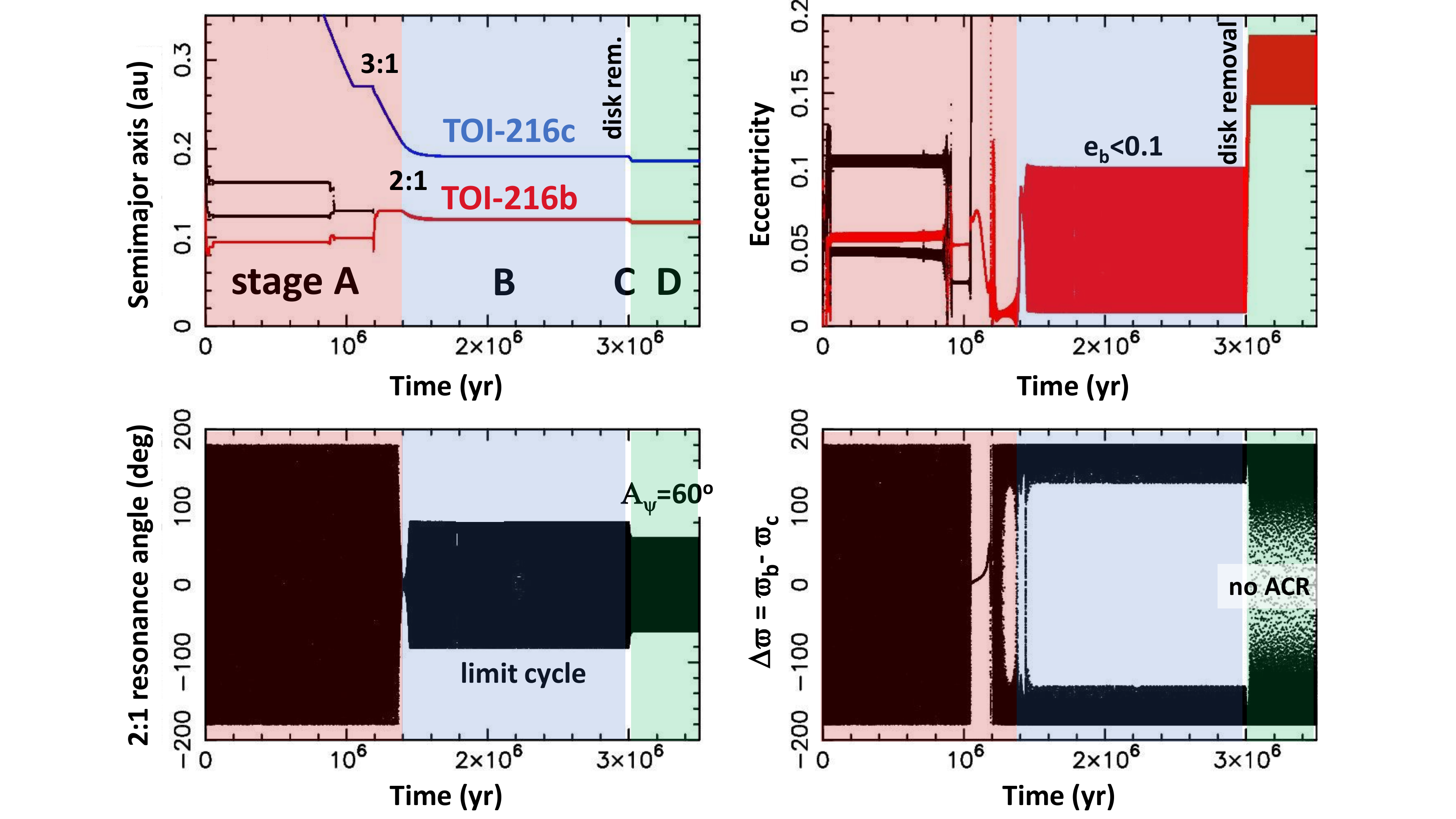}
\caption{A dynamical model with three inner super-Earths. Three planets, each with $m_{\rm p} = m_{\rm b}/3 \simeq 6.3$ 
$M_{\rm Earth}$, were initially placed at 0.15, 0.3 and 0.4 au and migrated inward. This initial stage is difficult 
to see in the figure because the migration is relatively fast. By $t=50,000$ yr after the start of the simulation 
the three planets settle in a stable resonant chain. The chain is destabilized as TOI-216c migrates in.
First, the two middle planets collide near $t=0.87$ Myr. Then, shortly after $t = 1$ Myr, TOI-216c is captured 
into the 3:1 resonance with the middle planet. The resonance is short-lived as overstable librations set in. The 
two inner planets are subsequently extracted from the resonance and merge, producing a final planet with the 
mass equal to that of TOI-216b. The following evolution is similar to that shown in Fig. 
\ref{model1}. The bottom panels show angular variables for the innermost planet and TOI-216c.} 
\label{model2}
\end{figure}

\begin{figure}
\epsscale{1.0}
\plotone{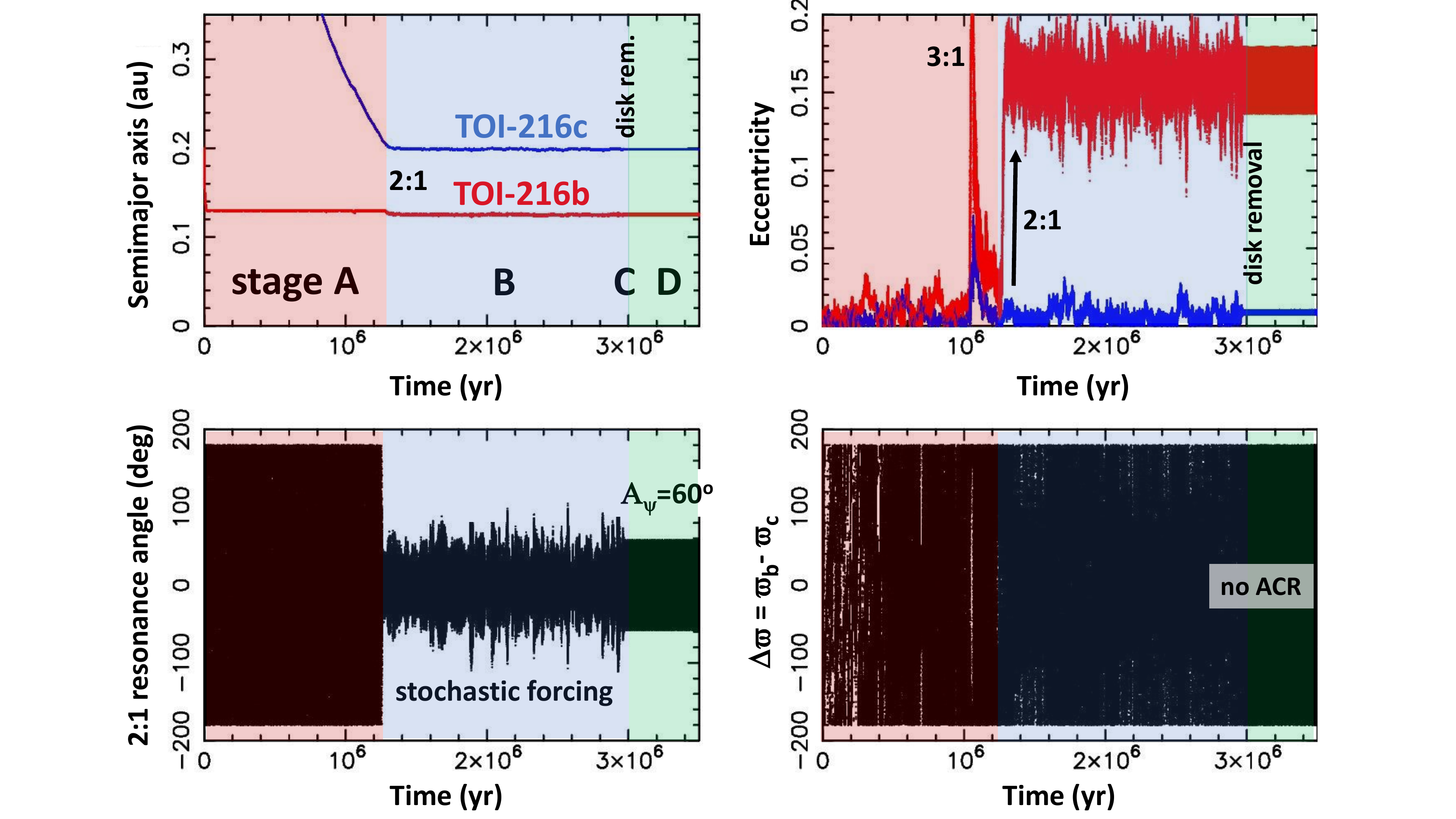}
\caption{A dynamical model with strong turbulent stirring. TOI-216b (red line in panel A) is held at the zero-torque 
radius and TOI-216c migrates in (blue line). The two planets become locked in the 2:1 resonance at time $t \simeq 1.25$ 
Myr after the start of the simulation. The resonant amplitude increases to $>40^\circ$ due to strong turbulent stirring
applied to both planets (Sect. 10). The gas disk is instantaneously removed at $t=3$ Myr; a more gradual disk removal 
would not change things as long as the stirring-to-damping strength remains the same (see discussion in the main text). 
The orbits of TOI-216b and c after the gas disk dispersal ($t>3$ Myr) are a good match to observations. } 
\label{model4}
\end{figure}

Some eccentricities could have been excited by the gravitational interaction between inner planets (Sect. 9 and Fig. 
\ref{model2}), but this would require that the eccentricity of TOI-216b was excited shortly before capture of 
TOI-216b and c into the 2:1 resonance; 
it would otherwise be damped to very low values (see Fig. \ref{model2}). The $\tau_e/\tau_a \sim h^2$ scaling 
strictly applies only to small planets that do not open gaps in the gas disk, but TOI-216b is expected 
to open a gap (Sect. 3). The timescale of eccentricity damping for the gap-opening planets is not well understood. 
It is possible that damping is not as strong as in the Type-I migration regime. Once $e>h$, the eccentricity damping 
is expected to weaken as $\tau_e \propto e^2$ (Papaloizou \& Larwood 2000, Cresswell et al. 2007), which may be 
consistent with the constraint obtained here.            

So far we discussed the cases with $p=0$ and $p=1$, with the former value being favored in the model with strong 
stochastic forcing (Sect. 10) and the latter value in the model with overstable librations (Sects. 8 and 9). 
By testing $0<p<1$ we found that a wide range of $p$-values works to generate the type of evolution shown in Figs. 
\ref{model1} and \ref{model2}, given that the $\tau_e/\tau_a$ ratio is slightly adjusted in each case. But there are 
limits to that. For example, we find that $p \lesssim 0.05$ is needed for stable librations with $e_{\rm eq} 
\simeq 0.16$, which would indicate a very weak coupling between $e$ and $a$ (Sect. 5). Although analytic limits from 
Goldreich \& Schlichting (2014) suggest that the limit cycles with $A_\psi \simeq 60^\circ$ should exist for these 
low values of $p$ (Sect. 7), we do not find them numerically: the libration amplitude stays zero for $p<0.05$ and grows 
beyond limits for $p>0.05$. Also, given the problems with the permanent capture of TOI-216b and c in the 3:1 resonance, 
we find that these very small values of $p$ probably do not apply (except, perhaps, in the case with a very strong 
stochastic forcing; Sect. 10). For comparison, Tanaka \& Ward (2004) found $p \simeq 0.4$ from their 3D disk model 
($e < h$ and no gap opening). If these results are applicable to TOI-216, this would favor the model with overstable 
librations and the fast inner disk removal (Sects. 8 and 9).

We adopted a number of simplifications in this work. The migration and damping torques, and their dependence
on radius, as described in Sect. 8, were kept fixed for the whole integration interval (except for the final 
stages when the torques were removed). In reality, the torques should 
have been modified as the disk's surface density, scale height and viscosity changed. This should not be a problem, 
however, as long as the TOI-216b,c evolved onto a limit cycle with $A_\psi \simeq 80^\circ$ during the late stages. 
For example, if $\tau_e$ and $\tau_a$ were shorter initially, the two planets would have more rapidly moved into 
the resonance, and if $\tau_e/\tau_a \simeq 0.02$, they would have more rapidly reached the limit cycle. 
The $\tau_e/\tau_a$ ratio should have changed as well. The cases with two planets 
in the 2:1 resonance, $\tau_e/\tau_a > 0.02$ and $p=1$ would not work, because the resonance would be short-lived
(Sect. 7). The cases with $\tau_e/\tau_a < 0.02$ would work as long as $\tau_e/\tau_a \rightarrow 0.02$ 
toward the end of the disk lifetime. If so, TOI-216b's eccentricity and the libration amplitude would 
more gradually increase during the disk lifetime. Similar considerations apply to the stochastic stirring model 
as well.

We adopted the MREF disk from Flock et al. (2019) and assumed that the zero-torque radius was at $\simeq 0.13$ 
au during the whole disk lifetime. This is unlikely to be realistic because the zero-torque radius is expected to 
move as the disk parameters change (Flock et al. 2016, 2019; Ueda et al. 2017). During the early disk stages, when 
the accretion rate and surface density was high, the MRI/DZ transition was likely located at larger orbital 
radii (due to the viscous heating; Schobert et al. 2019). If TOI-216b,c formed and migrated early, they 
could have been captured in the 2:1 resonance early as well with the two planets way out beyond their present 
orbital radii. They would migrate in, as the disk cools down, with TOI-216b following the zero-torque radius. 
The effects of viscous heating decrease in importance for $\dot{M}<10^{-8}$ $M_{\odot}$ yr$^{-1}$ (Schobert et al. 
2019) and can be neglected for $\dot{M} \lesssim 3 \times 10^{-9}$ $M_{\odot}$ yr$^{-1}$ (Flock et al. 2019). Thus, 
the situation should gradually approach the conditions investigated here (as $\dot{M}$ decreased).
Additional complications could arise from planetary mass loss/gain (Matsumoto \& Ogihara 2020), tides (e.g., Lithwick \& 
Wu 2012, Batygin \& Morbidelli 2013), etc., but these effects do not seem important a priori for TOI-216.

\begin{figure}
\epsscale{0.49}
\plotone{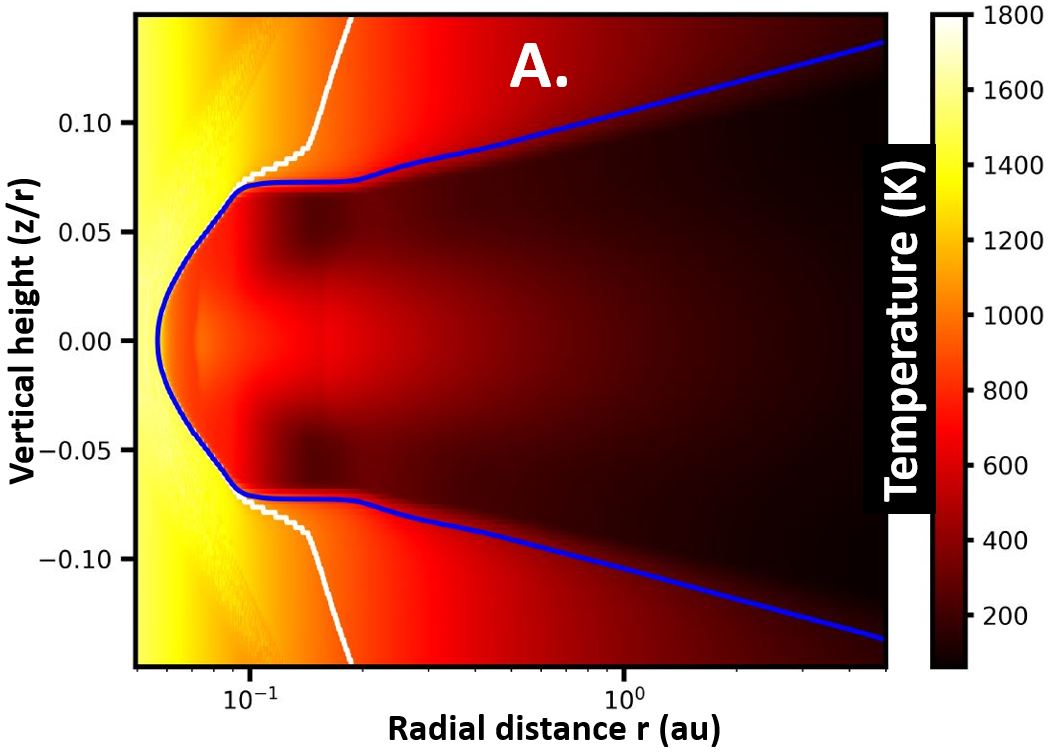}
\epsscale{0.48}
\plotone{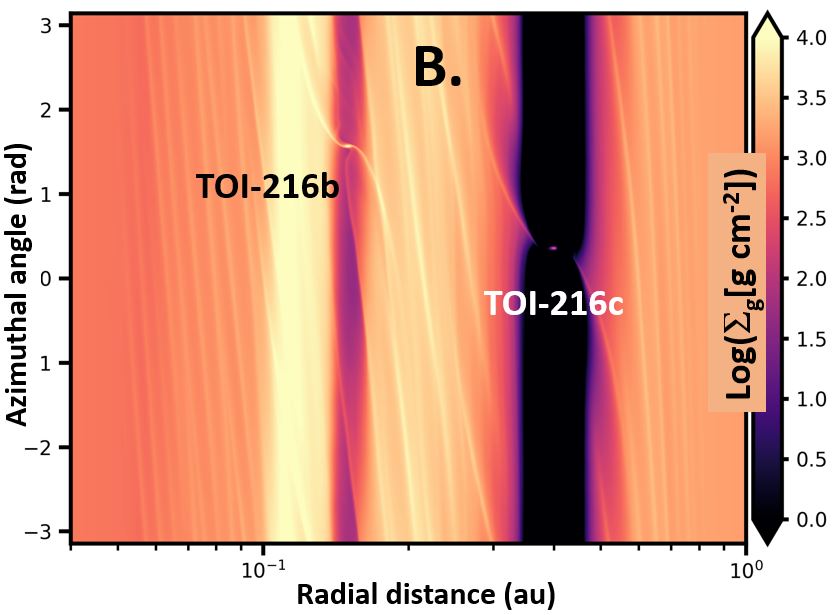}
\vspace*{-3mm}
\caption{A. The vertical structure of the inner protoplanetary disk obtained with \texttt{Fargo3D}
(Ben\'{\i}tez-Llambay \& Masset 2016). In this preliminary simulation, we reproduce the MREF disk structure
from Flock et al. (2019). To this end, we modified \texttt{Fargo3D} to solve the \textit{hydrostatic} 
radiation-hydrodynamics (RHD) equations in the radial and vertical directions (no azimuthal cells, no viscous 
heating). The disk viscosity was given by the usual $\alpha$ prescription (Shakura \& Sunyaev 1973) with 
$\alpha_{\rm MRI}=0.01$ in the inner MRI-active zone and $\alpha_{\rm DZ}=0.001$ in the outer dead zone. The $\alpha$ 
transition was set at $T_{\rm MRI}=900$ K (Desch \& Turner 2015), with the temperature being self-consistently 
computed in the code. The disk structure features a hot dust halo inside of the inner rim at 0.06~au, a curved 
dust rim between 0.06 and 0.1 au, a small shadowed region at 0.1-0.2~au, and a flared disk beyond 0.2 au. 
The white and blue lines in A are the silicate sublimation front and optical depth $\tau=2/3$ for the starlight 
irradiation. B. TOI-216b and c open gaps and migrate -- here in a locally isothermal 2D disk, $\alpha=0.01$ 
for $r<0.1$ au and $\alpha=0.001$ for $r>0.1$ au. The simulation with \textit{hydrodynamic} Fargo3D was run 
over $10^5$ periods of the inner planet.} 
\label{hydro}
\end{figure}

There is a number of potentially complex hydrodynamical effects that we did not take into account in this work. 
If TOI-216b opened a gap in the gas disk near the zero-torque radius, this would influence the surface density 
profile in the region where $\Sigma(r)$ rapidly changes with radius, and would feed back on torques experienced 
by TOI-216b. Would the planet trap at the MRI/DZ transition be strong enough to hold TOI-216b in place (Ataiee 
\& Kley 2021)? When TOI-216b evolved onto a large-eccentricity orbit near the inner disk's edge, it would 
move -- when following the orbit between the pericenter and apocenter -- in a radial interval where the disk 
properties abruptly change. How would the migration/damping work in this situation (e.g., Ogihara et al. 2010)? 
Moreover, TOI-216c could have opened a very wide gap which, at least in some situations, could have reached all 
the way down to TOI-216b (once the two planets moved onto the close-in orbits in the 2:1 resonance). Would this 
create favorable conditions for the formation of the present orbital architecture of TOI-216 planets, or could these 
cases be ruled out? Hydrodynamical studies will be needed to answer these questions, and this is left for future 
work (Fig. \ref{hydro}).  
  
We also need to clarify a few things about the constraint on the inner disk removal timescale, $\Delta t \sim 
25,000$ yr, in the model with overstable librations.
Strictly speaking this constraint is tied to the migration timescale of TOI-216c at the time of the gas disk 
dispersal, which depends on $\alpha$, $h$ and $\Sigma$ via Eq. (\ref{taua}). Here we opted for $\alpha=10^{-3}$, 
$h=0.02$ and $\Sigma$, as appropriate for the MREF disk model from Flock et al. (2019) ($\dot{M}=3 \pi \Sigma \nu 
= 3.6 \times 10^{-9}$ $M_{\odot}$ yr$^{-1}$), which gives $\tau_{a,2}\simeq0.8$ Myr (Sect. 3). The TOI-216 constraint 
therefore implies $\Delta t/\tau_{a,2} \sim 0.03$. So, for example, if the accretion rate (and surface density) was 10 
times lower at the time of the inner disk removal, which is a factor $\sim$25 lower than the MMSN (at 1 au), with 
everything else being the same, this would imply $\Delta t \sim 250,000$ yr. We quote an intermediate value, 
$\Delta t \sim 10^5$ yr, in the abstract. It is also possible that $\alpha$ in the dead zone was significantly lower 
than the one adopted here. This may formally not have a large effect on $\tau_{a,2}$, as $1/\tau_{a,2} \propto 
\alpha \Sigma$ (Eq. \ref{taua}) and $\Sigma \propto 1 / \alpha$; the $\alpha$ dependence thus cancels out for fixed 
$\dot{M}$. For fixed $\Sigma$ (e.g., the disk wind model; Suzuki et al. 2010, Bai \& Stone 2013, Simon et al. 
2013), $\tau_{a,2} \propto 1/\alpha$ instead. The migration of gap-opening planets in very low viscosity disks ($\alpha 
\lesssim 10^{-5}$) is complex and poorly understood (e.g., Lega et al. 2021).
 
In the model described in Sects. 8 and 9, the disk needs to be removed from inside out such as the migration/damping 
torques are first removed from TOI-216b and then, after $\Delta t$, from TOI-216c. The magnetospheric cavity expansion 
during the late stages of the disk evolution (e.g., Liu et al. 2017) could be responsible for this. Alternatively, 
TOI-216c could have carved a deep gap in the disk and cut the supply of gas from $>0.2$ au to $<0.2$ au. The inner 
disk would then be removed on the viscous timescale (Crida \& Morbidelli 2007)
\begin{equation}
\tau_\alpha =  { a^2 \over \alpha H^2 \Omega}\ . 
\end{equation}
For $a=0.12$ au, $\alpha = 0.001$, and $h = H/a = 0.02$, we obtain $\tau_\alpha \sim 40,000$ yr. This means that 
the inner disk could be removed very quickly. In summary, the reduction of the inflowing gas material from the 
outer disk regions by TOI-216c should lead to a relatively fast inner disk depletion (e.g., Tanigawa \& Tanaka 
2016). For things to work, however, the migration of TOI-216c would have to stop as well, and it is not clear 
whether this is the case because TOI-216c would still feel torques from the outer disk.

It has been established that the TOI-216 planets are not in the apsidal corotation resonance (Fig. \ref{real};
ACR, see Beaug\'e \& Ferraz-Mello 2003 and Hadden \& Payne 2020 for context). Instead, the periapse longitude 
difference, $\Delta \varpi = \varpi_{\rm b}-\varpi_{\rm c}$, circulates with the period of $P_{\Delta \varpi} \simeq 23$ yr. 
This may seem surprising because simple capture in the resonance should lead to the ACR locking. The examples of our model with overstable 
librations shown in Figs. \ref{model1} and \ref{model2} do not end up in ACR, but we were unable to tune 
things up well enough to replicate the observed behavior of $\Delta \varpi$. This happens because the 
eccentricity of TOI-216c ends up too low and $\varpi_2$ has fast and nonuniform precession. There is 
no source of excitation of $e_{\rm c}$ in this model other than the resonant interaction with TOI-216b. 
Here we neglected the apsidal precession of the two planets from the gas disk (e.g., Ali-Dib \& Petrovich 2020). We 
tested the influence of the disk potential using the scheme described in Vokrouhlick\'y \& Nesvorn\'y 
(2019) and found that it does not change things, because the inner disk in the MREF model has a relatively
low mass. The eccentricity of TOI-216c is excited in the model with the strong stochastic forcing 
(Fig. \ref{model4}). The final precession of $\varpi_1-\varpi_2$ in this model is a good match to 
observations.
  
\section{Conclusions}

The main results obtained in this work are 
\begin{enumerate}

\item TOI-216 is a system of two planets: the inner Neptune-class planet with the orbital period $P\simeq17$ d 
and outer half-Jupiter on a nearly circular orbit (Dawson et al. 2021). The observed TTVs are caused by the 
dynamical interaction of the two planets in the 2:1 resonance. The full TTV amplitudes, once the observations 
will cover the whole libration cycle, are expected to be $\sim 6,000$ min for TOI-216b 
(4.2 d or 24\% of the orbital period!)  and $\sim 1,000$ min for TOI-216c. 

\item We developed a dynamical model for capture of TOI-216 planets in the 2:1 resonance. To reach the resonance,
TOI-216b must have waited near the inner edge of the disk -- possibly near the zero torque radius at the 
transition from the outer dead to inner MRI-active zones (Flock et al. 2019). Alternatively, it could have 
been held in place by the one-sided torques at the magnetospheric cavity radius (e.g., Romanova et al. 2019).

\item Once TOI-216b's migration was stalled the two planets could 
have been captured into the 2:1 resonance. For the disk parameters adopted here, the capture is certain. 
With the two planets in the resonance, the main challenge is to explain the large eccentricity of TOI-216b 
($e_{\rm b}=0.16$) and large libration amplitude ($A_\psi \simeq 60^\circ$), without invoking special 
circumstances.

\item In the resonance, migrating TOI-216c pushed TOI-216b inside the zero-torque radius where the strong 
positive torques on TOI-216b compensated for TOI-216c's migration and held the two planets in place.
For $\tau_e/\tau_a \simeq 0.02$, the resonant interaction of planets would lead to a limit cycle with 
$e_1<0.1$ and $A_\psi\simeq80^\circ$ (for the nominal parametrization of torques with $p=1$; Sect. 5). 
The system could have remained in this configuration for the greater part of the protoplanetary disk lifetime. 

\item Then, if the inner disk was removed from inside out on a timescale $\Delta t \sim 0.03 \tau_{a,2}$ ($\sim 10^5$ yr
for the migration timescales of TOI-216c considered here), TOI-216b's eccentricity was boosted to the observed value, 
and $A_\psi$ slightly decreased (due to the adiabatic invariant conservation). The final configuration obtained in the 
model matches the observed orbital properties of the TOI-216 system. 

\item If there was a {\it Kepler}-like system of inner super-Earths/mini-Neptunes around TOI-216, it would have 
been destroyed by migrating TOI-216c. The inner planets would have merged into a single planet -- TOI-216b. 
The subsequent evolution of TOI-216b and c in the 2:1 resonance would be similar to that discussed for a pair 
of planets above.

\item The model with strong stochastic stirring is an interesting alternative to overstable librations. In this 
model, the overstable librations would need to be suppressed ($p \simeq 0$), and $\tau_e/\tau_a \simeq 0.05$. 
To obtain $A_\psi \simeq 60^\circ$, however, the stochastic diffusion would have to exceed, by at least two orders 
of magnitude, the one expected from the nominal turbulent stirring (Okuzumi \& Ormel 2013). This could be 
potentially accomplished if TOI-216b interacted with an inner vortex (e.g., Faure \& Nelson 2016),

\item Hydrodynamical simulations can be performed to test these models. The first question to ask is whether
$p \simeq 0$ (i.e., the semimajor axis migration and eccentricity damping are independent of each other; Eqs.
(9) and (10)), which would favor the stochastic stirring model, or $p > 0.05$, which would favor the model 
with overstable librations (for $p=1$ the eccentricity damping would happen at constant angular momentum; Sect. 5).   

\end{enumerate}

While interesting in its own right, the TOI-216 system may play an important role as a means of placing 
much-needed constraints on the nature of protoplanetary disks and their interaction with planets. Other 
resonant and near-resonant exoplanet pairs of interest include KOI-1599 (Panichi et al. 2019), Kepler-88 
(Nesvorn\'y et al. 2013, Barros et al. 2014), K2-19 (Petigura et al. 2020, Petit et al. 2020), K2-146 
(Hamann et al. 2019),  TOI-2202 (Trifonov et al. 2021), and HD 45364 (Hadden \& Payne 2020). The results 
obtained here can also be placed in the context of resonant and near-resonant planet chains, including
Kepler-11 (Lissauer et al. 2011, 2013), Kepler-60 (Go\'zdziewski et al. 2016), Kepler-80 (MacDonald et al. 2016), 
Kepler-223 (Mills et al. 2016), Kepler-444 (Campante et al. 2015, Papaloizou 2016, Mills \& Fabrycky 2017), 
K2-32 (Heller et al. 2019), K2-138 (Christiansen et al. 2018, Lopez et al. 2019), TOI-178  
(Leleu et al. 2021), and TRAPPIST-1 (Gillon et al. 2017, Luger et al. 2017, Ormel et al. 2017, Grimm et al. 
2018, Delrez et al. 2018, Agol et al. 2021), and close-in systems of {\it Kepler} planets in general
(Fabrycky et al. 2014, Winn \& Fabrycky 2015, Zhu \& Dong 2021).   

\acknowledgments
This work was supported by NASA's XRP program. The work of O.C. was supported by the Czech Science Foundation 
(grant 21-23067M), the Charles University Research program (No. UNCE/SCI/023), and the Ministry of 
Education, Youth and Sports of the Czech Republic through the e-INFRA CZ (ID:90140). 
M.F. acknowledges funding from the European Research Council (ERC) under the European Union’s Horizon 2020 
research and innovation program (grant agreement No. 757957). We thank Bekki Dawson for 
discussions.

\begin{table}
\centering
{
\begin{tabular}{lrr}
\hline \hline
Parameter & Value        &  Uncertainty   \\
\hline
\multicolumn{3}{c}{\textit{TOI-216}}      \\
$M_*\ (M_\odot)$ & 0.77      & 0.03         \\
$R_*\ (R_\odot)$ & 0.748     & 0.015       \\  
\hline
\multicolumn{3}{c}{\textit{Planet b}}      \\
$M_{\rm b}\ (M_{\rm Jup})$     & 0.059    & 0.002  \\ 
$P_{\rm b}$\ (day)           & 17.0968  & 0.0007 \\
$e_{\rm b}$                  & 0.160    & 0.003  \\
$\varpi_{\rm b}$ $(^\circ)$   & 291.8    & 1.0    \\
$\lambda_{\rm b}$ $(^\circ)$  & 82.5     & 0.3    \\
\hline
\multicolumn{3}{c}{\textit{Planet c}} \\
$M_{\rm c}\ (M_{\rm Jup})$     & 0.56     & 0.02  \\ 
$P_{\rm c}$\ (day)           & 34.5516  & 0.0003 \\
$e_{\rm c}$                  & 0.005   & 0.003  \\
$\varpi_{\rm c}$ $(^\circ)$   & 190      & 50    \\
$\lambda_{\rm c}$ $(^\circ)$  & 27.8     & 1.7   \\
\hline
\multicolumn{3}{c}{\textit{2:1 Resonance}} \\
$A_\psi$ $(^\circ)$         & 60       & 2     \\ 
$P_\psi$ (yr)              & 4        & --     \\
$P_{\Delta \varpi}$ (yr)      & 23        & --     \\ 
\hline
\end{tabular}
}
\caption{Parameters of the TOI-216 system from Dawson et al. (2021). Osculating orbital elements are given
at epoch BJD 2458325.3279. The three rows at the bottom are the resonant amplitude ($A_\psi$), period 
of resonant librations ($P_\psi$), and the circulation period of $\Delta \varpi = \varpi_b - \varpi_c$ 
($P_{\Delta \varpi}$). The periods were determined here from a numerical integration of the nominal solution 
(Fig. \ref{real}). \label{tab1}}
\end{table} 

\clearpage

\begin{thebibliography}

\bibitem[Adams et al.(2008)]{2008ApJ...683.1117A} Adams, F.~C., Laughlin, G., \& Bloch, A.~M.\ 2008, \apj, 683, 1117. doi:10.1086/589986

\bibitem[Agol et al.(2021)]{2021PSJ.....2....1A} Agol, E., Dorn, C., Grimm, S.~L., et al.\ 2021, PSJ, 2, 1. doi:10.3847/PSJ/abd022

\bibitem[Alexander et al.(2014)]{2014prpl.conf..475A} Alexander, R., Pascucci, I., Andrews, S., et al.\ 2014, Protostars and Planets VI, 475. doi:10.2458/azu\_uapress\_9780816531240-ch021

\bibitem[Ali-Dib \& Petrovich(2020)]{2020MNRAS.499..106A} Ali-Dib, M. \& Petrovich, C.\ 2020, \mnras, 499, 106. doi:10.1093/mnras/staa2820

\bibitem[Ataiee \& Kley(2021)]{2021A&A...648A..69A} Ataiee, S. \& Kley, W.\ 2021, \aap, 648, A69. doi:10.1051/0004-6361/202038772

\bibitem[Bai \& Stone(2013)]{2013ApJ...769...76B} Bai, X.-N. \& Stone, J.~M.\ 2013, \apj, 769, 76. doi:10.1088/0004-637X/769/1/76

\bibitem[Barros et al.(2014)]{2014A&A...561L...1B} Barros, S.~C.~C., D{\'\i}az, R.~F., Santerne, A., et al.\ 2014, \aap, 561, L1. doi:10.1051/0004-6361/201323067

\bibitem[Batygin(2015)]{2015MNRAS.451.2589B} Batygin, K.\ 2015, \mnras, 451, 2589. doi:10.1093/mnras/stv1063

\bibitem[Batygin \& Morbidelli(2013)]{2013AJ....145....1B} Batygin, K. \& Morbidelli, A.\ 2013, \aj, 145, 1. doi:10.1088/0004-6256/145/1/1

\bibitem[Batygin \& Adams(2017)]{2017AJ....153..120B} Batygin, K. \& Adams, F.~C.\ 2017, \aj, 153, 120. doi:10.3847/1538-3881/153/3/120

\bibitem[Batygin et al.(2015)]{2015AJ....149..167B} Batygin, K., Deck, K.~M., \& Holman, M.~J.\ 2015, \aj, 149, 167. doi:10.1088/0004-6256/149/5/167

\bibitem[Beauge \& Ferraz-Mello(1993)]{1993Icar..103..301B} Beaug\'e, C. \& Ferraz-Mello, S.\ 1993, Icarus, 103, 301. doi:10.1006/icar.1993.1072

\bibitem[Beauge \& Ferraz-Mello(1994)]{1994Icar..110..239B} Beaug\'e, C. \& Ferraz-Mello, S.\ 1994, Icarus, 110, 239. doi:10.1006/icar.1994.1119

\bibitem[Ben{\'\i}tez-Llambay \& Masset(2016)]{2016ApJS..223...11B} Ben{\'\i}tez-Llambay, P. \& Masset, F.~S.\ 2016, \apjs, 223, 11. doi:10.3847/0067-0049/223/1/11

\bibitem[Bouvier et al.(2014)]{2014prpl.conf..433B} Bouvier, J., Matt, S.~P., Mohanty, S., et al.\ 2014, Protostars and Planets VI, 433. doi:10.2458/azu\_uapress\_9780816531240-ch019

\bibitem[Campante et al.(2015)]{2015ApJ...799..170C} Campante, T.~L., Barclay, T., Swift, J.~J., et al.\ 2015, \apj, 799, 170. doi:10.1088/0004-637X/799/2/170

\bibitem[Chambers(1999)]{1999MNRAS.304..793C} Chambers, J.~E.\ 1999, \mnras, 304, 793. doi:10.1046/j.1365-8711.1999.02379.x

\bibitem[Chrenko \& Nesvorn{\'y}(2020)]{2020A&A...642A.219C} Chrenko, O. \& Nesvorn{\'y}, D.\ 2020, \aap, 642, A219. doi:10.1051/0004-6361/202038988

\bibitem[Christiansen et al.(2018)]{2018AJ....155...57C} Christiansen, J.~L., Crossfield, I.~J.~M., Barentsen, G., et al.\ 2018, \aj, 155, 57. doi:10.3847/1538-3881/aa9be0

\bibitem[Cimerman et al.(2018)]{2018A&A...618A.169C} Cimerman, N.~P., Kley, W., \& Kuiper, R.\ 2018, \aap, 618, A169. doi:10.1051/0004-6361/201833591

\bibitem[Cresswell et al.(2007)]{2007A&A...473..329C} Cresswell, P., Dirksen, G., Kley, W., et al.\ 2007, \aap, 473, 329. doi:10.1051/0004-6361:20077666

\bibitem[Crida et al.(2006)]{2006Icar..181..587C} Crida, A., Morbidelli, A., \& Masset, F.\ 2006, Icarus, 181, 587. doi:10.1016/j.icarus.2005.10.007

\bibitem[Crida et al.(2008)]{2008A&A...483..325C} Crida, A., S{\'a}ndor, Z., \& Kley, W.\ 2008, \aap, 483, 325. doi:10.1051/0004-6361:20079291

\bibitem[Crida \& Morbidelli(2007)]{2007MNRAS.377.1324C} Crida, A. \& Morbidelli, A.\ 2007, \mnras, 377, 1324. doi:10.1111/j.1365-2966.2007.11704.x

\bibitem[Dawson et al.(2019)]{2019AJ....158...65D} Dawson, R.~I., Huang, C.~X., Lissauer, J.~J., et al.\ 2019, \aj, 158, 65. doi:10.3847/1538-3881/ab24ba

\bibitem[Dawson et al.(2021)]{2021AJ....161..161D} Dawson, R.~I., Huang, C.~X., Brahm, R., et al.\ 2021, \aj, 161, 161. doi:10.3847/1538-3881/abd8d0

\bibitem[Deck \& Batygin(2015)]{2015ApJ...810..119D} Deck, K.~M. \& Batygin, K.\ 2015, \apj, 810, 119. doi:10.1088/0004-637X/810/2/119

\bibitem[Delisle et al.(2012)]{2012A&A...546A..71D} Delisle, J.-B., Laskar, J., Correia, A.~C.~M., et al.\ 2012, \aap, 546, A71. doi:10.1051/0004-6361/201220001

\bibitem[Delisle et al.(2015)]{2015A&A...579A.128D} Delisle, J.-B., Correia, A.~C.~M., \& Laskar, J.\ 2015, \aap, 579, A128. doi:10.1051/0004-6361/201526285

\bibitem[Delrez et al.(2018)]{2018MNRAS.475.3577D} Delrez, L., Gillon, M., Triaud, A.~H.~M.~J., et al.\ 2018, \mnras, 475, 3577. doi:10.1093/mnras/sty051

\bibitem[Dempsey \& Nelson(2018)]{2018ApJ...867...75D} Dempsey, A.~M. \& Nelson, B.~E.\ 2018, \apj, 867, 75. doi:10.3847/1538-4357/aae36c

\bibitem[Desch \& Turner(2015)]{2015ApJ...811..156D} Desch, S.~J. \& Turner, N.~J.\ 2015, \apj, 811, 156. doi:10.1088/0004-637X/811/2/156

\bibitem[Duffell \& MacFadyen(2013)]{2013ApJ...769...41D} Duffell, P.~C. \& MacFadyen, A.~I.\ 2013, \apj, 769, 41. doi:10.1088/0004-637X/769/1/41

\bibitem[Fabrycky et al.(2014)]{2014ApJ...790..146F} Fabrycky, D.~C., Lissauer, J.~J., Ragozzine, D., et al.\ 2014, \apj, 790, 146. doi:10.1088/0004-637X/790/2/146

\bibitem[Faure \& Nelson(2016)]{2016A&A...586A.105F} Faure, J. \& Nelson, R.~P.\ 2016, \aap, 586, A105. doi:10.1051/0004-6361/201527194

\bibitem[Flock et al.(2016)]{2016ApJ...827..144F} Flock, M., Fromang, S., Turner, N.~J., et al.\ 2016, \apj, 827, 144. doi:10.3847/0004-637X/827/2/144

\bibitem[Flock et al.(2017)]{2017ApJ...850..131F} Flock, M., Nelson, R.~P., Turner, N.~J., et al.\ 2017, \apj, 850, 131. doi:10.3847/1538-4357/aa943f

\bibitem[Flock et al.(2019)]{2019A&A...630A.147F} Flock, M., Turner, N.~J., Mulders, G.~D., et al.\ 2019, \aap, 630, A147. doi:10.1051/0004-6361/201935806

\bibitem[Frank et al.(1992)]{1992apa..book.....F} Frank, J., King, A., \& Raine, D.\ 1992, Camb. Astrophys. Ser., Vol. 21,

\bibitem[Freudenthal et al.(2018)]{2018A&A...618A..41F} Freudenthal, J., von Essen, C., Dreizler, S., et al.\ 2018, \aap, 618, A41. doi:10.1051/0004-6361/201833436

\bibitem[Fung et al.(2014)]{2014ApJ...782...88F} Fung, J., Shi, J.-M., \& Chiang, E.\ 2014, \apj, 782, 88. doi:10.1088/0004-637X/782/2/88

\bibitem[Gillon et al.(2017)]{2017Natur.542..456G} Gillon, M., Triaud, A.~H.~M.~J., Demory, B.-O., et al.\ 2017, \nat, 542, 456. doi:10.1038/nature21360

\bibitem[Goldreich \& Schlichting(2014)]{2014AJ....147...32G} Goldreich, P. \& Schlichting, H.~E.\ 2014, \aj, 147, 32. doi:10.1088/0004-6256/147/2/32

\bibitem[Go{\'z}dziewski et al.(2016)]{2016MNRAS.455L.104G} Go{\'z}dziewski, K., Migaszewski, C., Panichi, F., et al.\ 2016, \mnras, 455, L104. doi:10.1093/mnrasl/slv156

\bibitem[Gomes(1995)]{1995Icar..115...47G} Gomes, R.~S.\ 1995, Icarus, 115, 47. doi:10.1006/icar.1995.1077

\bibitem[Grimm et al.(2018)]{2018A&A...613A..68G} Grimm, S.~L., Demory, B.-O., Gillon, M., et al.\ 2018, \aap, 613, A68. doi:10.1051/0004-6361/201732233

\bibitem[Hadden \& Payne(2020)]{2020AJ....160..106H} Hadden, S. \& Payne, M.~J.\ 2020, \aj, 160, 106. doi:10.3847/1538-3881/aba751

\bibitem[Hamann et al.(2019)]{2019AJ....158..133H} Hamann, A., Montet, B.~T., Fabrycky, D.~C., et al.\ 2019, \aj, 158, 133. doi:10.3847/1538-3881/ab32e3

\bibitem[Hayashi(1981)]{1981PThPS..70...35H} Hayashi, C.\ 1981, Progress of Theoretical Physics Supplement, 70, 35. doi:10.1143/PTPS.70.35

\bibitem[Heller et al.(2019)]{2019A&A...625A..31H} Heller, R., Rodenbeck, K., \& Hippke, M.\ 2019, \aap, 625, A31. doi:10.1051/0004-6361/201935276

\bibitem[Henrard \& Lemaitre(1983)]{1983CeMec..30..197H} Henrard, J. \& Lemaitre, A.\ 1983, Celestial Mechanics, 30, 197. doi:10.1007/BF01234306

\bibitem[Holman et al.(2010)]{2010Sci...330...51H} Holman, M.~J., Fabrycky, D.~C., Ragozzine, D., et al.\ 2010, Science, 330, 51. doi:10.1126/science.1195778

\bibitem[Isella \& Natta(2005)]{2005A&A...438..899I} Isella, A. \& Natta, A.\ 2005, \aap, 438, 899. doi:10.1051/0004-6361:20052773

\bibitem[Kanagawa \& Szuszkiewicz(2020)]{2020ApJ...894...59K} Kanagawa, K.~D. \& Szuszkiewicz, E.\ 2020, \apj, 894, 59. doi:10.3847/1538-4357/ab862f

\bibitem[Kanagawa et al.(2018)]{2018ApJ...861..140K} Kanagawa, K.~D., Tanaka, H., \& Szuszkiewicz, E.\ 2018, \apj, 861, 140. doi:10.3847/1538-4357/aac8d9

\bibitem[Kipping et al.(2019)]{2019MNRAS.486.4980K} Kipping, D., Nesvorn{\'y}, D., Hartman, J., et al.\ 2019, \mnras, 486, 4980. doi:10.1093/mnras/stz1141

\bibitem[Kretke \& Lin(2012)]{2012ApJ...755...74K} Kretke, K.~A. \& Lin, D.~N.~C.\ 2012, \apj, 755, 74. doi:10.1088/0004-637X/755/1/74

\bibitem[Laughlin et al.(2004)]{2004ApJ...608..489L} Laughlin, G., Steinacker, A., \& Adams, F.~C.\ 2004, \apj, 608, 489. doi:10.1086/386316

\bibitem[Laughlin et al.(2005)]{2005ApJ...622.1182L} Laughlin, G., Butler, R.~P., Fischer, D.~A., et al.\ 2005, \apj, 622, 1182. doi:10.1086/424686

\bibitem[Lee \& Peale(2002)]{2002ApJ...567..596L} Lee, M.~H. \& Peale, S.~J.\ 2002, \apj, 567, 596. doi:10.1086/338504

\bibitem[Lega et al.(2021)]{2021A&A...646A.166L} Lega, E., Nelson, R.~P., Morbidelli, A., et al.\ 2021, \aap, 646, A166. doi:10.1051/0004-6361/202039520

\bibitem[Leleu et al.(2021)]{2021A&A...649A..26L} Leleu, A., Alibert, Y., Hara, N.~C., et al.\ 2021, \aap, 649, A26. doi:10.1051/0004-6361/202039767

\bibitem[Levison \& Duncan(1994)]{1994Icar..108...18L} Levison, H.~F. \& Duncan, M.~J.\ 1994, Icarus, 108, 18. doi:10.1006/icar.1994.1039

\bibitem[Lissauer et al.(2011)]{2011Natur.470...53L} Lissauer, J.~J., Fabrycky, D.~C., Ford, E.~B., et al.\ 2011, \nat, 470, 53. doi:10.1038/nature09760

\bibitem[Lissauer et al.(2013)]{2013ApJ...770..131L} Lissauer, J.~J., Jontof-Hutter, D., Rowe, J.~F., et al.\ 2013, \apj, 770, 131. doi:10.1088/0004-637X/770/2/131

\bibitem[Lithwick \& Wu(2012)]{2012ApJ...756L..11L} Lithwick, Y. \& Wu, Y.\ 2012, \apjl, 756, L11. doi:10.1088/2041-8205/756/1/L11

\bibitem[Liu et al.(2017)]{2017A&A...601A..15L} Liu, B., Ormel, C.~W., \& Lin, D.~N.~C.\ 2017, \aap, 601, A15. doi:10.1051/0004-6361/201630017

\bibitem[Lopez et al.(2019)]{2019A&A...631A..90L} Lopez, T.~A., Barros, S.~C.~C., Santerne, A., et al.\ 2019, \aap, 631, A90. doi:10.1051/0004-6361/201936267

\bibitem[Luger et al.(2017)]{2017NatAs...1E.129L} Luger, R., Sestovic, M., Kruse, E., et al.\ 2017, Nature Astronomy, 1, 0129. doi:10.1038/s41550-017-0129

\bibitem[MacDonald et al.(2016)]{2016AJ....152..105M} MacDonald, M.~G., Ragozzine, D., Fabrycky, D.~C., et al.\ 2016, \aj, 152, 105. doi:10.3847/0004-6256/152/4/105

\bibitem[Malhotra(1995)]{1995AJ....110..420M} Malhotra, R.\ 1995, \aj, 110, 420. doi:10.1086/117532

\bibitem[Malygin et al.(2014)]{2014A&A...568A..91M} Malygin, M.~G., Kuiper, R., Klahr, H., et al.\ 2014, \aap, 568, A91. doi:10.1051/0004-6361/201423768

\bibitem[Marcy et al.(2001)]{2001ApJ...556..296M} Marcy, G.~W., Butler, R.~P., Fischer, D., et al.\ 2001, \apj, 556, 296. doi:10.1086/321552

\bibitem[Mart{\'\i} et al.(2013)]{2013MNRAS.433..928M} Mart\'{\i}, J.~G., Giuppone, C.~A., \& Beaug{\'e}, C.\ 2013, \mnras, 433, 928. doi:10.1093/mnras/stt765

\bibitem[Masset et al.(2006)]{2006ApJ...642..478M} Masset, F.~S., Morbidelli, A., Crida, A., et al.\ 2006, \apj, 642, 478. doi:10.1086/500967

\bibitem[Matsumoto \& Ogihara(2020)]{2020ApJ...893...43M} Matsumoto, Y. \& Ogihara, M.\ 2020, \apj, 893, 43. doi:10.3847/1538-4357/ab7cd7

\bibitem[Meyer \& Wisdom(2008)]{2008Icar..193..213M} Meyer, J. \& Wisdom, J.\ 2008, carus, 193, 213. doi:10.1016/j.icarus.2007.09.008

\bibitem[Millholland et al.(2018)]{2018AJ....155..106M} Millholland, S., Laughlin, G., Teske, J., et al.\ 2018, \aj, 155, 106. doi:10.3847/1538-3881/aaa894

\bibitem[Mills \& Fabrycky(2017)]{2017ApJ...838L..11M} Mills, S.~M. \& Fabrycky, D.~C.\ 2017, \apjl, 838, L11. doi:10.3847/2041-8213/aa6543

\bibitem[Mills et al.(2016)]{2016Natur.533..509M} Mills, S.~M., Fabrycky, D.~C., Migaszewski, C., et al.\ 2016, \nat, 533, 509. doi:10.1038/nature17445

\bibitem[Nelson(2005)]{2005A&A...443.1067N} Nelson, R.~P.\ 2005, \aap, 443, 1067. doi:10.1051/0004-6361:20042605

\bibitem[Nelson et al.(2016)]{2016MNRAS.455.2484N} Nelson, B.~E., Robertson, P.~M., Payne, M.~J., et al.\ 2016, \mnras, 455, 2484. doi:10.1093/mnras/stv2367

\bibitem[Nesvorn{\'y} \& Vokrouhlick{\'y}(2016)]{2016ApJ...823...72N} Nesvorn{\'y}, D. \& Vokrouhlick{\'y}, D.\ 2016, \apj, 823, 72. doi:10.3847/0004-637X/823/2/72

\bibitem[Nesvorn{\'y} et al.(2013)]{2013ApJ...777....3N} Nesvorn{\'y}, D., Kipping, D., Terrell, D., et al.\ 2013, \apj, 777, 3. doi:10.1088/0004-637X/777/1/3

\bibitem[Ogihara et al.(2010)]{2010ApJ...721.1184O} Ogihara, M., Duncan, M.~J., \& Ida, S.\ 2010, \apj, 721, 1184. doi:10.1088/0004-637X/721/2/1184

\bibitem[Okuzumi \& Ormel(2013)]{2013ApJ...771...43O} Okuzumi, S. \& Ormel, C.~W.\ 2013, \apj, 771, 43. doi:10.1088/0004-637X/771/1/43

\bibitem[Ormel et al.(2017)]{2017A&A...604A...1O} Ormel, C.~W., Liu, B., \& Schoonenberg, D.\ 2017, \aap, 604, A1. doi:10.1051/0004-6361/201730826

\bibitem[Paardekooper et al.(2010)]{2010MNRAS.401.1950P} Paardekooper, S.-J., Baruteau, C., Crida, A., et al.\ 2010, \mnras, 401, 1950. doi:10.1111/j.1365-2966.2009.15782.x

\bibitem[Paardekooper et al.(2013)]{2013MNRAS.434.3018P} Paardekooper, S.-J., Rein, H., \& Kley, W.\ 2013, \mnras, 434, 3018. doi:10.1093/mnras/stt1224

\bibitem[Panichi et al.(2019)]{2019MNRAS.485.4601P} Panichi, F., Migaszewski, C., \& Go{\'z}dziewski, K.\ 2019, \mnras, 485, 4601. doi:10.1093/mnras/stz721

\bibitem[Papaloizou(2016)]{2016CeMDA.126..157P} Papaloizou, J.~C.~B.\ 2016, Celestial Mechanics and Dynamical Astronomy, 126, 157. doi:10.1007/s10569-016-9689-9

\bibitem[Papaloizou \& Larwood(2000)]{2000MNRAS.315..823P} Papaloizou, J.~C.~B. \& Larwood, J.~D.\ 2000, \mnras, 315, 823. doi:10.1046/j.1365-8711.2000.03466.x

\bibitem[Petigura et al.(2020)]{2020AJ....159....2P} Petigura, E.~A., Livingston, J., Batygin, K., et al.\ 2020, \aj, 159, 2. doi:10.3847/1538-3881/ab5220

\bibitem[Petit et al.(2020)]{2020MNRAS.496.3101P} Petit, A.~C., Petigura, E.~A., Davies, M.~B., et al.\ 2020, \mnras, 496, 3101. doi:10.1093/mnras/staa1736

\bibitem[Rein(2012)]{2012MNRAS.427L..21R} Rein, H.\ 2012, \mnras, 427, L21. doi:10.1111/j.1745-3933.2012.01337.x

\bibitem[Rein \& Papaloizou(2009)]{2009A&A...497..595R} Rein, H. \& Papaloizou, J.~C.~B.\ 2009, \aap, 497, 595. doi:10.1051/0004-6361/200811330



\bibitem[Rivera et al.(2005)]{2005ApJ...634..625R} Rivera, E.~J., Lissauer, J.~J., Butler, R.~P., et al.\ 2005, \apj, 634, 625. doi:10.1086/491669

\bibitem[Rivera et al.(2010)]{2010ApJ...719..890R} Rivera, E.~J., Laughlin, G., Butler, R.~P., et al.\ 2010, \apj, 719, 890. doi:10.1088/0004-637X/719/1/890

\bibitem[Robert et al.(2018)]{2018A&A...617A..98R} Robert, C.~M.~T., Crida, A., Lega, E., et al.\ 2018, \aap, 617, A98. doi:10.1051/0004-6361/201833539

\bibitem[Romanova et al.(2019)]{2019MNRAS.485.2666R} Romanova, M.~M., Lii, P.~S., Koldoba, A.~V., et al.\ 2019, \mnras, 485, 2666. doi:10.1093/mnras/stz535

\bibitem[Schobert et al.(2019)]{2019ApJ...881...56S} Schobert, B.~N., Peeters, A.~G., \& Rath, F.\ 2019, \apj, 881, 56. doi:10.3847/1538-4357/ab2df6

\bibitem[Shakura \& Sunyaev(1973)]{1973A&A....24..337S} Shakura, N.~I. \& Sunyaev, R.~A.\ 1973, \aap, 500, 33

\bibitem[Sidlichovsky \& Nesvorny(1994)]{1994A&A...289..972S} Sidlichovsky, M. \& Nesvorny, D.\ 1994, \aap, 289, 972

\bibitem[Simon et al.(2013)]{2013ApJ...775...73S} Simon, J.~B., Bai, X.-N., Armitage, P.~J., et al.\ 2013, \apj, 775, 73. doi:10.1088/0004-637X/775/1/73

\bibitem[Snellgrove et al.(2001)]{2001A&A...374.1092S} Snellgrove, M.~D., Papaloizou, J.~C.~B., \& Nelson, R.~P.\ 2001, \aap, 374, 1092. doi:10.1051/0004-6361:20010779

\bibitem[Suzuki et al.(2010)]{2010ApJ...718.1289S} Suzuki, T.~K., Muto, T., \& Inutsuka, S.-. ichiro .\ 2010, \apj, 718, 1289. doi:10.1088/0004-637X/718/2/1289

\bibitem[Tanaka \& Ward(2004)]{2004ApJ...602..388T} Tanaka, H. \& Ward, W.~R.\ 2004, \apj, 602, 388. doi:10.1086/380992

\bibitem[Tanigawa \& Tanaka(2016)]{2016ApJ...823...48T} Tanigawa, T. \& Tanaka, H.\ 2016, \apj, 823, 48. doi:10.3847/0004-637X/823/1/48

\bibitem[Terquem \& Papaloizou(2019)]{2019MNRAS.482..530T} Terquem, C. \& Papaloizou, J.~C.~B.\ 2019, \mnras, 482, 530. doi:10.1093/mnras/sty2693

\bibitem[Trifonov et al.(2021)]{2021arXiv210805323T} Trifonov, T., Brahm, R., Espinoza, N., et al.\ 2021, arXiv:2108.05323

\bibitem[Ueda et al.(2017)]{2017ApJ...843...49U} Ueda, T., Okuzumi, S., \& Flock, M.\ 2017, \apj, 843, 49. doi:10.3847/1538-4357/aa74b5

\bibitem[Vokrouhlick{\'y} \& Nesvorn{\'y}(2019)]{2019CeMDA.132....3V} Vokrouhlick{\'y}, D. \& Nesvorn{\'y}, D.\ 2019, Celestial Mechanics and Dynamical Astronomy, 132, 3. doi:10.1007/s10569-019-9941-1

\bibitem[Williams \& Cieza(2011)]{2011ARA&A..49...67W} Williams, J.~P. \& Cieza, L.~A.\ 2011, \araa, 49, 67. doi:10.1146/annurev-astro-081710-102548

\bibitem[Winn \& Fabrycky(2015)]{2015ARA&A..53..409W} Winn, J.~N. \& Fabrycky, D.~C.\ 2015, \araa, 53, 409. doi:10.1146/annurev-astro-082214-122246

\bibitem[Wolf \& Voshchinnikov(2004)]{2004CoPhC.162..113W} Wolf, S. \& Voshchinnikov, N.~V.\ 2004, Computer Physics Communications, 162, 113. doi:10.1016/j.cpc.2004.06.070

\bibitem[Youdin \& Goodman(2005)]{2005ApJ...620..459Y} Youdin, A.~N. \& Goodman, J.\ 2005, \apj, 620, 459. doi:10.1086/426895

\bibitem[Zhu \& Dong(2021)]{2021arXiv210302127Z} Zhu, W. \& Dong, S.\ 2021, arXiv:2103.02127

\end{thebibliography}
\end{document}